\documentclass{aa}  
\usepackage{graphicx}
\usepackage{txfonts}

\newcommand{\control}{{\tt\string Full Dataset}}
\newcommand{\nobal}{{\tt\string Non-BAL Only }}

\def\heii{He\,{\sc ii}~}
\def\mgii{Mg\,{\sc ii}~}

\def\civ{C\,{\sc iv}~}
\def\ciii{C\,{\sc iii}~}
\def\siiv{Si\,{\sc iv}~}
\def\aliii{Al\,{\sc iii}~}

\def\feii{Fe\,{\sc ii}~}

\def\kms{km~s$^{-1}$}
\usepackage{csquotes}
\usepackage{lineno}
\usepackage{float}

\usepackage{array}      
\usepackage{lipsum}
\begin{document}

   \title{Spectroscopic Quasar Anomaly Detection  (SQuAD)}

   \subtitle{I: Rest-Frame UV Spectra from SDSS DR16}

   \author{Arihant Tiwari
          \inst{1,2}
          \and
          M. Vivek\inst{2}
          }

   \institute{Indian Institute of Science Education and Research, Bhopal, Madhya Pradesh 462066, India\\
              \email{arihant19@iiserb.ac.in}
         \and
             Indian Institute of Astrophysics, 100 Feet Rd, Santhosapuram, 2\textsuperscript{nd} Block, Koramangala, Bengaluru, Karnataka, India\\
             \email{vivek.m@iiap.res.in}
             }

   \date{Received XXXX, 2024; accepted XXXX, 2024}


  \abstract
   {}
   {We present the results of applying anomaly detection algorithms to a quasar spectroscopic sub-sample from the SDSS DR16 Quasar Catalog, covering the redshift range  1.88 $\leq$ z $\leq$ 2.47.}
   {Principal Component Analysis (PCA) was employed for dimensionality reduction of the quasar spectra, followed by hierarchical K-Means clustering in a 20-dimensional PCA eigenvector hyperspace. To prevent broad absorption line (BAL) quasars from being identified as the primary anomaly group, we conducted the analysis with and without them, comparing both datasets for a clearer identification of other anomalous quasar types.}
   {We identified 1,888 anomalous quasars, categorized into 10 broad groups. The anomalous groups include \civ Peakers—quasars with extremely strong and narrow \civ emission lines; Excess \siiv emitters—quasars where the \siiv line is as strong as the \civ line; and \siiv Deficient anomalies, which exhibit significantly weaker \siiv emission compared to typical quasars.The anomalous nature of these quasars is attributed to lower Eddington ratios for \civ Peakers, super-solar metallicity for Excess \siiv emitters, and sub-solar metallicity for \siiv Deficient anomalies. Additionally, we identified four groups of BAL anomalies: Blue BALs, Flat BALs, Reddened BALs, and FeLoBALs, distinguished primarily by the strength of reddening in these sources. Further, among the non-BAL quasars, we identified three types of reddened anomaly groups classified as heavily reddened, moderately reddened, and plateau-shaped spectrum quasars, each exhibiting varying degrees of reddening. The detected anomalies are presented as a value-added catalog.
}
   {}


   \maketitle
%

\section{Introduction} \label{sec:intro}

Quasars/ Active Galactic Nuclei (AGNs), are highly luminous galactic cores powered by gas accreting onto a supermassive black hole, emitting radiation across the entire electromagnetic spectrum \citep{1973A&A....24..337S}. AGNs emit radiation through various physical mechanisms contributing to their broadband spectrum. Thermal emission from the accretion disk includes X-rays from hot inner regions and optical/UV light from cooler outer areas. Synchrotron radiation, produced by relativistic electrons spiraling in magnetic fields, is prominent in radio-loud AGNs and spans from radio waves to X-rays. Bremsstrahlung radiation occurs when electrons decelerate near ions, emitting X-rays significant in hot gas regions like the corona or narrow-line region. Inverse Compton scattering in the corona boosts low-energy photons to higher energies via interactions with high-energy electrons, contributing to X-ray and gamma-ray emissions \citep[for e.g., see,][for a review]{Begelman1984,haardt1991,koratkar1999}.

The UV spectrum of an AGN often follows a power-law distribution, expressed as $F_\nu \propto \nu^{-\alpha}$, where the spectral index, $\alpha$ typically ranges from 0.5 to 1.5. The power-law continuum arises from the superposition of multiple blackbody spectra emitted by the accretion disk's varying radial temperature zones. 
Additionally, the UV emission exhibits strong emission and absorption lines with broad and narrow components, often modeled using double/single Gaussian profiles. These lines involve ions such as \civ, \siiv, \aliii, \mgii,  and \heii. They serve as key diagnostics of the physical conditions in the AGN's environment \citep[for e.g.,][]{zheng1997composite,brotherton2001composite,2016AJ....151..155H}.

Due to the multiple physical processes involved, quasars exhibit a wide range of properties that can differ greatly from one quasar to another, or even over time within the same quasar  \citep[for e.g.,][]{wilhite2005,2012MNRAS.421L.107V,2012MNRAS.423.2879V,lamassa2015,green2022time}.  In UV spectra, features such as the spectral slope and the strength of emission lines display significant variability among quasars.  Processes like gravitational lensing and obscuration by gas clouds along the line of sight also contribute to this spectral variability. \citep[e.g.][]{wiklind1996redshift}. 

Several past studies have revealed the presence of a variety of unusual quasar spectra that deviate significantly from the standard SED such as the \citet{2001AJ....122..549V} composite, and have confirmed the existence of unusual or anomalous quasar types for e.g. \citet{plotkin2008large} present a sample of unusual BL Lac objects analytically selected from SDSS DR5, \citet{diamond2009high} identify a sample of 74 high-redshift quasars ($z>3$) with weak emission lines from the SDSS DR5 and \citet{hines2001hubble} analyze two unusually spectro-polarized QSOs. \citet{artymowicz1993metallicity} study the AGN metallicity as a function of unusual star formation activities at high redshifts. Two \enquote{extraordinarily red} quasars are found and their dust reddening is characterized by \citet{gregg2002reddest} from the Faint Images of the Radio Sky at Twenty-centimeters (FIRST) survey. \citet{collin2002quasars} study the disc dynamics and supermassive blackhole (SMBH) mass of a quasar population with super-Eddington accretion. \citet{hall2002unusual} identified 23 unusual broad absorption line (BAL) quasars with diverse properties, including a quasar with the most recorded absorption lines and others with complex \mgii\ absorption patterns. Additionally, they found rare cases of low-ionization broad absorption line (LoBAL) quasars and iron low-ionization broad absorption-line (FeLoBAL) quasars with strong Fe II emissions, and unique cases of Fe III absorption without Fe II. 
Studying unusual quasar spectra is important as it can reveal physical processes or environmental factors that are not apparent in typical quasars. These unusual characteristics may result from enhanced mechanisms such as star formation, accretion activity, reprocessed emission, outflows or feedback. By examining these outliers, we can gain valuable insights into the diversity of quasar properties and improve our understanding of the underlying physics governing quasars.

With the massive number of quasars being discovered every day, it is easy to miss out on an extremely interesting or rare object that might get covered in the pile of data being pumped by survey telescopes all around the world. Machine learning serves as a golden tool in this scenario as it can be used to quickly pick out anomalous or weird-behaving objects (quasars in this case) from all of the \enquote{normal} samples. Several studies have focused on detecting peculiar objects from large datasets, with some specifically using spectra for anomaly detection. For example, \citet{meusinger2012unusual} identified around 1,000 spectroscopically unusual quasars in SDSS DR7 using Kohonen Maps, while \citet{REIS2021100437} evaluated the effectiveness of various outlier detection algorithms applied to SDSS galactic spectra. Similarly, \citet{refId0} extracted a sample of spectroscopic anomalies from the All-Sky Wide-field Infrared Survey Explorer (AllWISE) Sky Survey.

The high dimensionality and complexity of astronomical spectral data hinder the effective application of anomaly detection algorithms. In such cases, Principal Component Analysis (PCA) offers a robust method for dimensionality reduction. This technique transforms the original spectral features into a lower-dimensional space, retaining the most critical variance in the data. The efficiency of PCA in preprocessing data for clustering algorithms like K-Means is well-established. PCA enhances the clustering process by reducing noise and irrelevant variations, allowing K-Means to operate more effectively on transformed data. \citet{Xiong2018} and \citet{henrion2013casos} illustrate this by adopting PCA for anomaly detection in large-scale astronomical datasets. They emphasize that the use of PCA not only simplifies the data but also improves the performance of clustering algorithms by focusing on the most informative aspects of the spectra.

In this study, we apply a hierarchical K-Means clustering approach to detect and analyze spectroscopic anomalies in quasars from the Sloan Digital Sky Survey: Sixteenth Data Release Quasar (SDSS DR16Q) Catalog \citep{Lyke_2020}. We chose the K-Means algorithm due to its interpretability, which makes it easier to identify which characteristics of the data contribute to cluster formation. While more sophisticated algorithms might offer marginal improvements in clustering accuracy, they are often computationally expensive and more challenging to interpret, making K-Means a balanced choice for efficient and meaningful analysis. K-Means works effectively when combined with PCA, which reduces the dimensionality of the data, keeping only the most important variance. By leveraging PCA for dimensionality reduction, we distill essential features from the high-dimensional spectral data, enhancing K-Means' ability to identify anomalous quasar spectra. The clustering process groups similar spectra, allowing for the detection of outliers within each cluster, which are considered anomalous. Given the large sample size, hierarchical clustering further refines the data into subgroups with distinct and consistent property trends, enabling a more detailed analysis of quasar spectral anomalies.  

The Spectroscopic Quasar Anomaly Detection (SQuAD) project aims to detect spectroscopically anomalous quasars from large-scale spectroscopic surveys and conduct follow-up studies on the most interesting cases. This paper is the first in a series, with a focus specifically on identifying these anomalies in the rest-frame UV spectra from the SDSS DR16Q catalog. The paper is organized as follows: In \S \ref{sec:Data}, we describe the sample selection, data reduction, cleaning procedures, and dimensionality reduction using PCA. \S \ref{sec:Clustering} covers our clustering techniques, including K-Means and its hierarchical application, along with methods for determining the optimal number of clusters. \S \ref{sec:Results} presents our main findings, with visual representations of the detected anomalies. In \S \ref{sec:Discussion}, we categorize the identified anomalies and discuss the observable properties of each group, as well as the implications for the physical properties of the quasars. Finally, \S \ref{sec:Conclusion} summarizes the findings.

\section{Data} \label{sec:Data}
The spectral data for this project is obtained from the SDSS DR16Q catalog  which contains 750,414 quasars in the redshift range from $0 < z \leq 7.1$. The spectra were obtained using the SDSS and  Baryon Acoustic Oscillation Spectroscopic Survey (BOSS)  spectrographs from the SDSS survey, covering wavelengths from 3600 to 10,400 Å (with good throughput between 3650 and 9500 Å) and a resolution of 1560–2270 in the blue channel and 1850–2650 in the red channel \citep{smee2013multi}.

\begin{figure}[h!]
   \centering
   \includegraphics[width=\hsize]{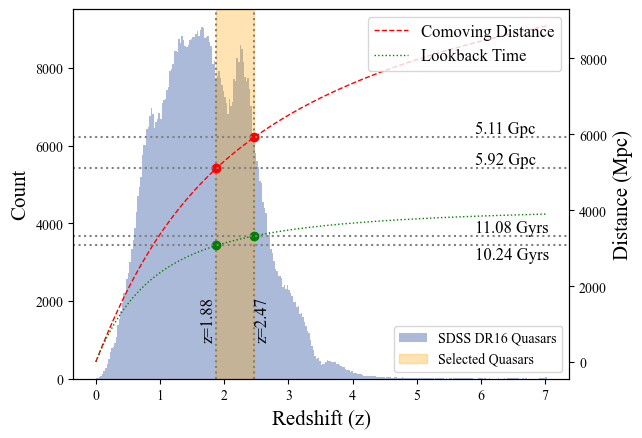}
      \caption{Redshift distribution of quasars in the SDSS DR16 catalog (blue) and the selected subset of quasars (yellow) within the specified redshift range, $1.88 \leq z \leq 2.47$. The red dashed curve represents the comoving distance, while the green dotted curve shows the lookback time as a function of redshift. The corresponding distance and time values at two key redshifts: $z = 1.88$ and $z = 2.47$, are marked.}
    \label{fig:redshift_dist}
   \end{figure}
   
Our goal is to identify anomalous quasar spectra in this dataset. To facilitate outlier detection, we defined a \enquote{similar-looking} sample space by restricting the rest-frame wavelength range of all  spectra to 1250–3000 \AA. This range captures four prominent emission lines typical of quasar spectra:  \siiv $\lambda 1400$\AA, \civ $\lambda 1549$\AA, \ciii $\lambda 1909$\AA, and \mgii $\lambda 2798$\AA\, \citep[see][]{2001AJ....122..549V}. With the BOSS spectrograph's wavelength coverage of $[3600,10400]\AA$ , our wavelength window translated to a redshift range of $z\in[1.88,2.47]$ as shown in Fig.~\ref{fig:redshift_dist}, where the yellow shaded region marks the selected sample.

Initially, our analysis included all quasars within the specified redshift range. However, this led to a systematic issue where broad absorption line (BAL) quasars were predominantly identified as anomalies, as discussed  in \S \ref{sec:Discussion}. To address this, we conducted our analysis using two nested datasets: a) {\tt\string Full Dataset} comprising all 81,814 confirmed quasars in the sample (comprising $\sim13\%$ of the DR16Q population), and b) {\tt\string Non-BAL Only Dataset}, a subset of the {\tt\string Full Dataset}, containing 55,245 non-BAL quasars. Note that the {\tt\string Full Dataset} includes 26,569 BAL quasars, resulting in a BAL fraction of 32.47\% for this dataset. The BAL quasars were selected based on the keyword \enquote{BAL PROB} value being greater than 0.5 \citep{Guo_2019}. Additionally, a signal-to-noise ratio (SNR) of $\geq5$, value of the \enquote{IS\_QSO\_FINAL} parameter = 1 and no ZWARNING flag was required for all included quasar spectra.

\subsection{Data Preprocessing} \label{subsec:Data Preprocessing}

The quasar spectra from SDSS contain observed flux $(ergs/cm^2/sec/\mathrm{\AA})$ distributed logarithmically on a wavelength $(\mathrm{\AA})$ scale. Given the wide range of redshift, we first corrected all spectra to the rest frame using the equation: $\lambda_0 = \frac{\lambda_{obs}}{z+1}$ -- where $\lambda_0$ is the rest wavelength, $\lambda_{obs}$ is the observed wavelength, and $z$ is the quasar redshift. The rest frame spectra underwent a four step pre-processing procedure (to remove noise and artifacts and prepare them for the clustering algorithm), which included the following steps chronologically:

\begin{enumerate}

    \item \textbf{Normalization:} \label{process:Normalization}The spectra were normalized using the maximum flux value \citep[e.g.][]{liu2011method}, as described in Equation \ref{eq: Normalization}, thereby adjusting the flux value range for all spectra to [-1,1]. This normalization reduced the likelihood of the algorithm identifying a spectrum as an outlier based on extreme luminosity, whether high or low.
    
    \begin{center}
    \centering
        \begin{equation}\label{eq: Normalization}
            (F_{normalized})_\lambda = \frac{F_\lambda}{\textbf{max}(\{F_\lambda\})}
        \end{equation}
    \end{center}

    \item \textbf{Smoothing}: The normalized spectra were smoothed using a Savitzky–Golay filter \citep{1964AnaCh..36.1627S}, with a window length of five pixels and a third-degree polynomial. This method effectively reduced noise without altering the length or overall shape of the spectral array. However, it did not remove artifacts like extremely narrow spikes in the spectra, which are often caused by cosmic ray hits or system-induced errors, as discussed by \citet{newman2004mass}.

    \item \textbf{Re-Sampling: } The smoothed, normalized spectra were resampled following the method of \citet{carnall2017spectres} to bring all spectra onto a common wavelength grid, specifically 1250 to 3000 \AA. This resampling also reduced the data size by a factor of 2. A typical SDSS spectrum, originally with a flux array length of around 4000, was reduced to approximately 850 points after resampling with 2 \AA\ binning. This significantly lowered computational costs without compromising the quality of the information retained by the spectra.

    \item \textbf{Padding: } As previously discussed, all spectra were shifted to their rest wavelengths and resampled onto a common grid spanning 1250 \AA\ to 3000 \AA\ for standardization.  Due to the varying redshift values $(z)$, not all spectra naturally cover this entire wavelength range. Consequently, any flux outside this range was excluded, and gaps at both ends were filled by padding with the trailing flux value. This procedure ensured that all spectra had uniform dimensions, which is essential for the anomaly detection algorithms. 
    
\end{enumerate}
\begin{figure}[h!]
   \centering
   \includegraphics[width=\hsize]{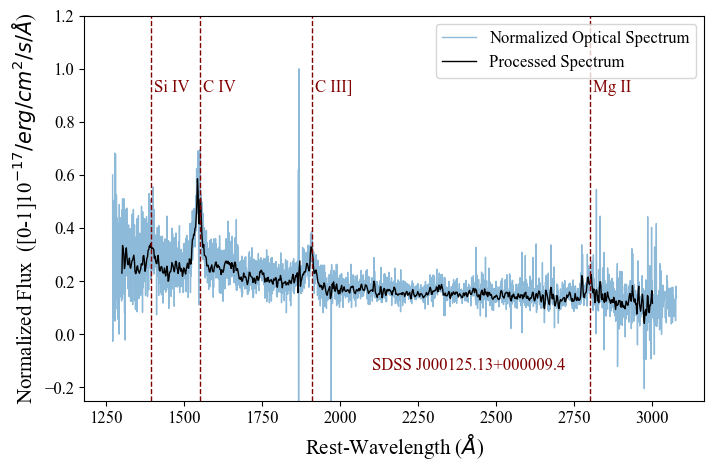}
    \caption{An example quasar spectrum showing the max-normalized flux before (blue) and after pre-processing (black). Note that the pre-processing effectively reduces noise without altering the overall spectral shape.}
    \label{fig:Processed Spectrum}
   \end{figure}

We initially attempted to utilize the {\tt\string AND} and {\tt\string OR} masks to obtain a clean spectrum. However, these methods proved ineffective as they removed large portions of the spectra, leading to detection as anomalies due to the missing chunks. Instead, a simple Savitzky-Golay (Sav-Gol) smoothing and re-binning approach, as discussed earlier, was more effective in eliminating bad pixels and producing a cleaner spectrum. 

Similarly, we also tried using median normalization, median filtering to remove cosmic rays, and various rebinning methods but could not match the performance of the currently adopted methodology. For example, median normalization reduced the explained PCA variance to 75\% instead of the current 92\%. The large noise in the continuum, especially in the low SNR spectra, distributes the variance across a larger number of components, reducing the variance captured by the first few principal components in PCA.
Fig.~\ref{fig:Processed Spectrum} shows a typical spectrum before (in blue) and after pre-processing (black). 

\subsection{Principal Component Analysis}\label{subsec: PCA}

As discussed by \citet{2010ApJ...714..487S}, a spectrum can be represented as  a vector in a high-dimensional space with as many unit vectors as the wavelength grid and the flux values being the coefficients for these unit vectors.
\begin{figure}[h!]
    \centering
    \includegraphics[width=\hsize]{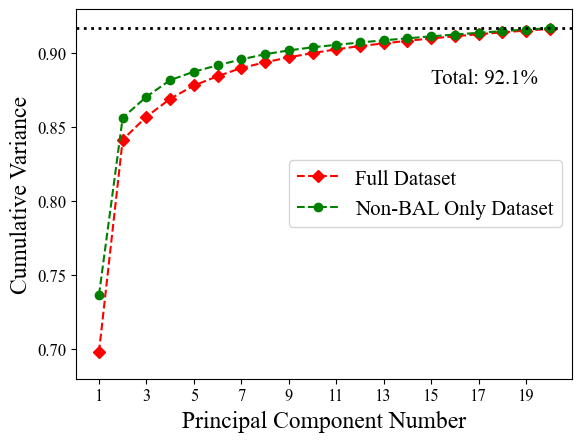}
    \caption{The figure shows the cumulative variance as a function of the PCA components for the PCA decomposition with 20 components, for both the {\tt \string Full} (red) and {\tt \string Non-BAL Only} (green) datasets. In both cases, a total explained variance of 92.1\% is achieved with 20 PCA components.}
    \label{fig:PCA_Explained_Variance}
\end{figure}
Therefore, the quasar spectral catalog can be considered a set of vectors in this space with well defined (Euclidean) distances between each vector pair. Since the performance of K-Means clustering is inversely proportional to the number of dimensions in which clustering is performed \citep[e.g.][]{10.5120/1789-2471}, we apply PCA to reduce the dimensions of the spectra from $\sim$ 850-dimensional wavelength-flux hyperspace to a 20 dimension PCA eigenvector hyperspace. The number of components was chosen such that we obtain a cumulative explained variance greater than 90\%, which was achieved  with 20 PCA components. This accounted for 92.1\% of the total variance, as depicted in Fig.~\ref{fig:PCA_Explained_Variance}. Fig.~\ref{fig:pca_eigenspectra} in appendix, shows the 20 eigenvectors from the PCA decomposition, which represent the principal components capturing the most significant variance in the dataset. These components are crucial for understanding the underlying structure of the quasar spectra, as they highlight key recurring patterns. Among the 20 PCA components, it is noteworthy that the second PCA eigenvector captures the reddening component of a given spectra as evident from its steep positively sloped shape (see component 2, Fig.~\ref{fig:pca_eigenspectra}). Therefore, if a quasar has a high coefficient for the second PCA eigenvector (PCA 2), it indicates a greater amount of reddening. In other words, the PCA 2 coefficient is directly proportional to the level of reddening in the spectrum. To address the remaining 8\% of the variance not captured by the PCA model, we reconstruct each spectrum using the 20 PCA eigenvectors and compute the residual error using Equation \ref{Eq: Reconstruction Error} as:

\begin{equation}\label{Eq: Reconstruction Error}
    \text{Residual Error} = \frac{\Sigma_i^N\sqrt{(F_O|\lambda_i - F_R|\lambda_i)^2}}{N}
\end{equation}
\begin{quote}
    Here, $F_O|\lambda_i$ denotes the original flux at a given wavelength $(\lambda_i)$, $F_R$ denotes the reconstructed flux value at the same wavelength, and $N$ is the total number of wavelength points.
\end{quote}

\begin{figure}[h!]
    \centering
    \includegraphics[width=\hsize]{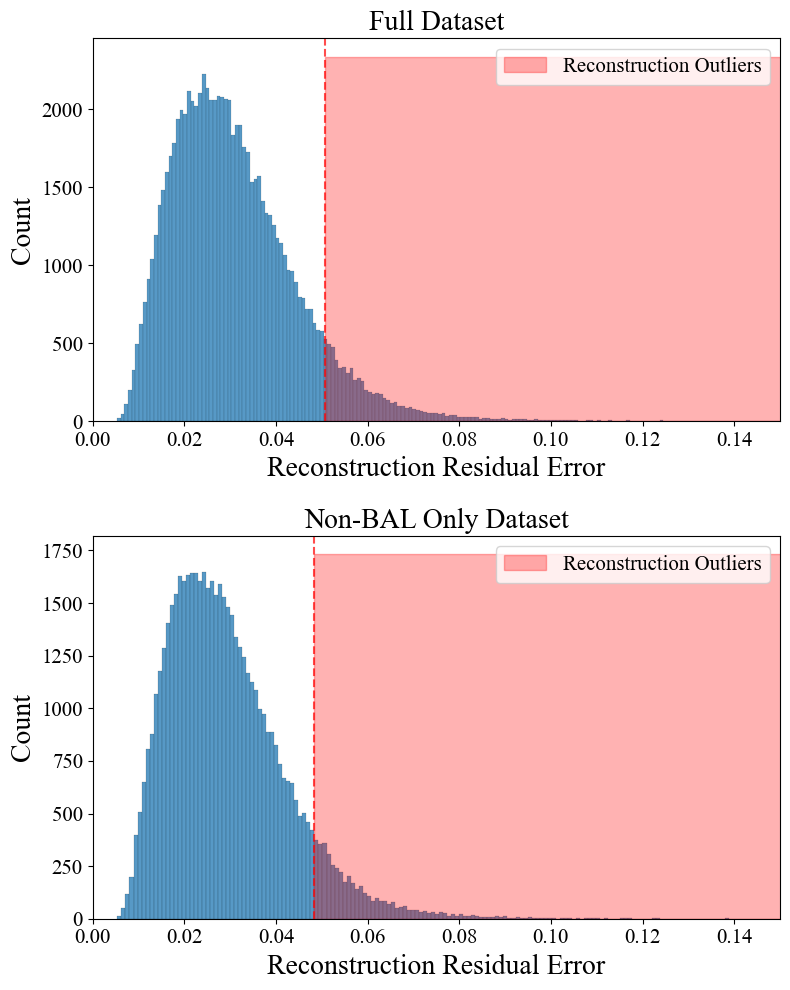}
    \caption{The distribution of residual error for PCA spectral reconstruction is shown for both the {\tt\string Full} (top) and {\tt\string Non-BAL Only} (bottom) datasets. Quasars within the red shaded region are classified as \enquote{reconstruction outliers} and are excluded from the main datasets prior to clustering.}
    \label{fig:Reconstruction_Error_Distribution}
\end{figure}

The top 8\textsuperscript{th} percentile of spectra in the residual error distribution, shown as the red shaded region in Fig.~\ref{fig:Reconstruction_Error_Distribution} 
represents those spectra insufficiently explained by the PCA model. These spectra, referred to as \enquote{Reconstruction Outliers} are removed from the main sample but are handled separately, as discussed in section~\ref{subsec:ReconstructionOutlierGrouping}. This ensures that the clustering algorithm operates only on well-explained spectra.

\subsection{Reconstruction Outlier Grouping }
\label{subsec:ReconstructionOutlierGrouping}
Quasars with large reconstruction errors naturally qualify as \enquote{anomalous} quasars. However, including these spectra in the main dataset would require a larger number of clusters. Most of the reconstruction outliers consisted of quasars with large flat sections, typically introduced during padding to compensate for missing data. These flat sections resulted in extended chunks of missing values. To address this, a simple algorithm was used to flag and remove spectra containing such large flat regions, ensuring that these reconstruction outliers were systematically excluded from the analysis. To the remaining spectra, we applied K-Means clustering, as described in \S\ref{sec:Clustering}.  The algorithm grouped them into three categories: Red BAL quasars, FeLoBALs, and heavily reddened quasars (see \S \ref{sec:Discussion} for more details about each). These groups corresponded to anomaly categories that emerged from the clustering of the main sample and were subsequently assigned to their appropriate classifications.  

\section{Clustering} \label{sec:Clustering}
\begin{figure*}[h!]
   \centering
   \includegraphics[width=\hsize]{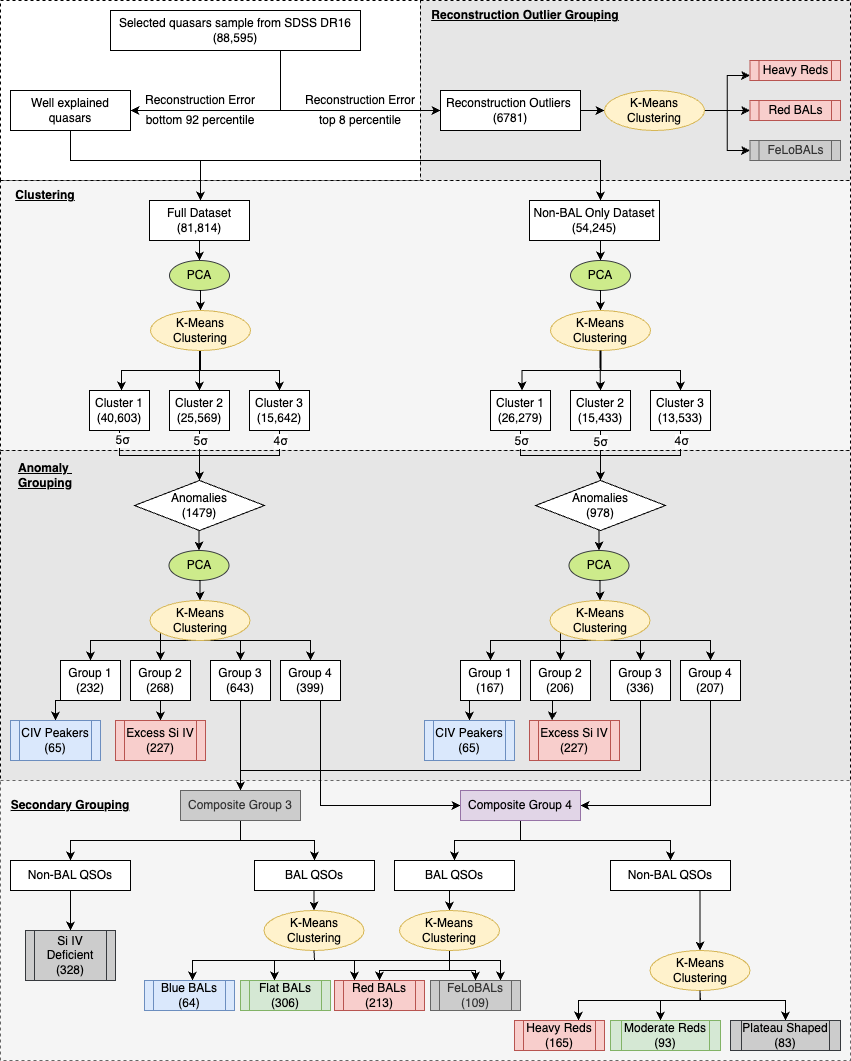}

   \caption{The flow chart shows the  steps followed by our algorithm, beginning  from the quasar sample selection to the final anomaly groups obtained.}
    \label{fig:flowchart}
    \end{figure*}
\begin{figure}[h!]
    \centering
    \includegraphics[width=\hsize]{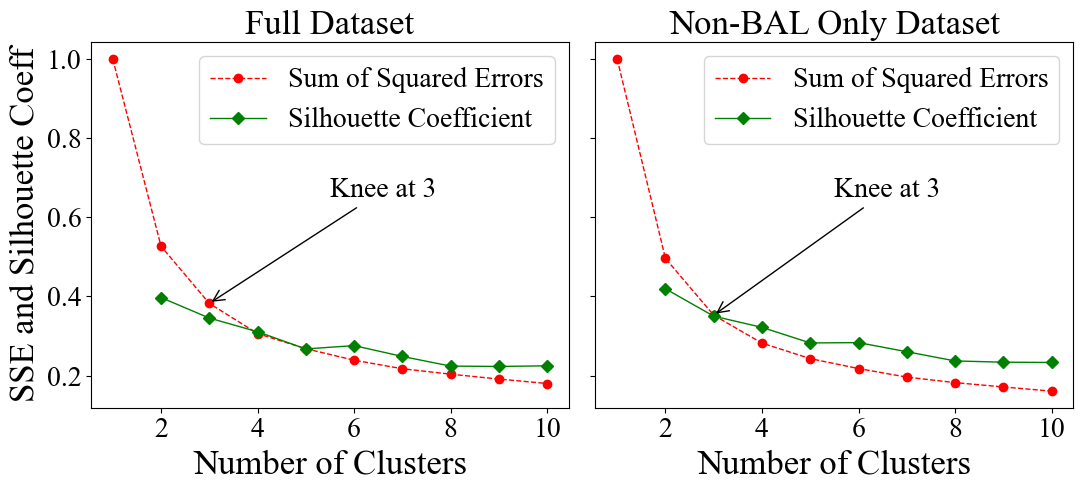}
    \caption{The figure shows the Sum of Squared Errors (SSE) and Silhouette Coefficients as a function of cluster numbers for the Full (left) and Non-BAL Only (right) datasets. The optimal number of clusters is determined using both the elbow (knee) method and the silhouette coefficient for both datasets. The knee of the SSE curve occurs at 3 for both datasets.}
    \label{fig:Optimum_Clusters}
\end{figure}

K-Means clustering \citep[e.g.][]{Bradley1998RefiningIP} was employed to categorize quasars based on the Euclidean distances between their eigenvectors in a 20-dimensional PCA coefficient space. The algorithm initiates by selecting $k$ random eigenspectra from the dataset, assuming these to be the centroids of distinct clusters. Subsequently, all spectra in the sample are assigned to the nearest centroid according to Euclidean distance. The cluster centroids are then recalculated as the mean of the spectra within each cluster. This iterative process continues until the cluster memberships stabilize, with no spectra re-assigned in two consecutive iterations. The final output comprises $k$ cluster centroids and the classification of each spectrum into one of the $k$ clusters. K-Means is well-suited for large datasets due to its simplicity and efficiency. However, the algorithm requires the user to input a predefined number of clusters, $k$. Instead of assuming this number based on prior knowledge, we calculated it explicitly using the elbow method on the Sum of Squared Errors (SSE) and Silhouette Coefficients \citep[e.g.][]{syakur2018integration}, as shown in  Fig.~\ref{fig:Optimum_Clusters}. The \enquote{knee} or elbow was found using the Kneed Python module which utilizes the Kneedle algorithm \citep{satopaa2011finding} to find the point of maximum curvature, which in a well behaved clustering problem, represents the optimum number of clusters for the distribution. The elbow method determines the optimal number of clusters by identifying the point of maximum curvature in the SSE plot. This "elbow" indicates where increasing the number of clusters yields minimal improvements in reducing error. Based on this analysis, $k=3$ is obtained as the optimum number of clusters for both of our datasets. Detailed cluster visualizations are presented in \S \ref{subsec: Results_Clusters}. 
The number of quasars present in each cluster is given in Table \ref{table:Quasars per cluster}:

\begin{table}[h!]
\begin{tabular}{|c | c | c | c|} 
 \hline
 Dataset & Cluster 1 & Cluster 2 & Cluster 3 \\ [0.5ex] 
 \hline
 {\tt\string Full} & 40,603 & 25,569 & 15,642 \\ 
 {\tt\string Non-BAL} & 26,279 & 15,433 & 13,533 \\
 \hline
\end{tabular}
\caption{Number of quasars in each cluster for both datasets}
\label{table:Quasars per cluster}
\end{table}

\subsection{Anomaly Detection} \label{subsec: Anomaly Detection}
\begin{figure*}[h!]
    \centering
    \includegraphics[width=\hsize]{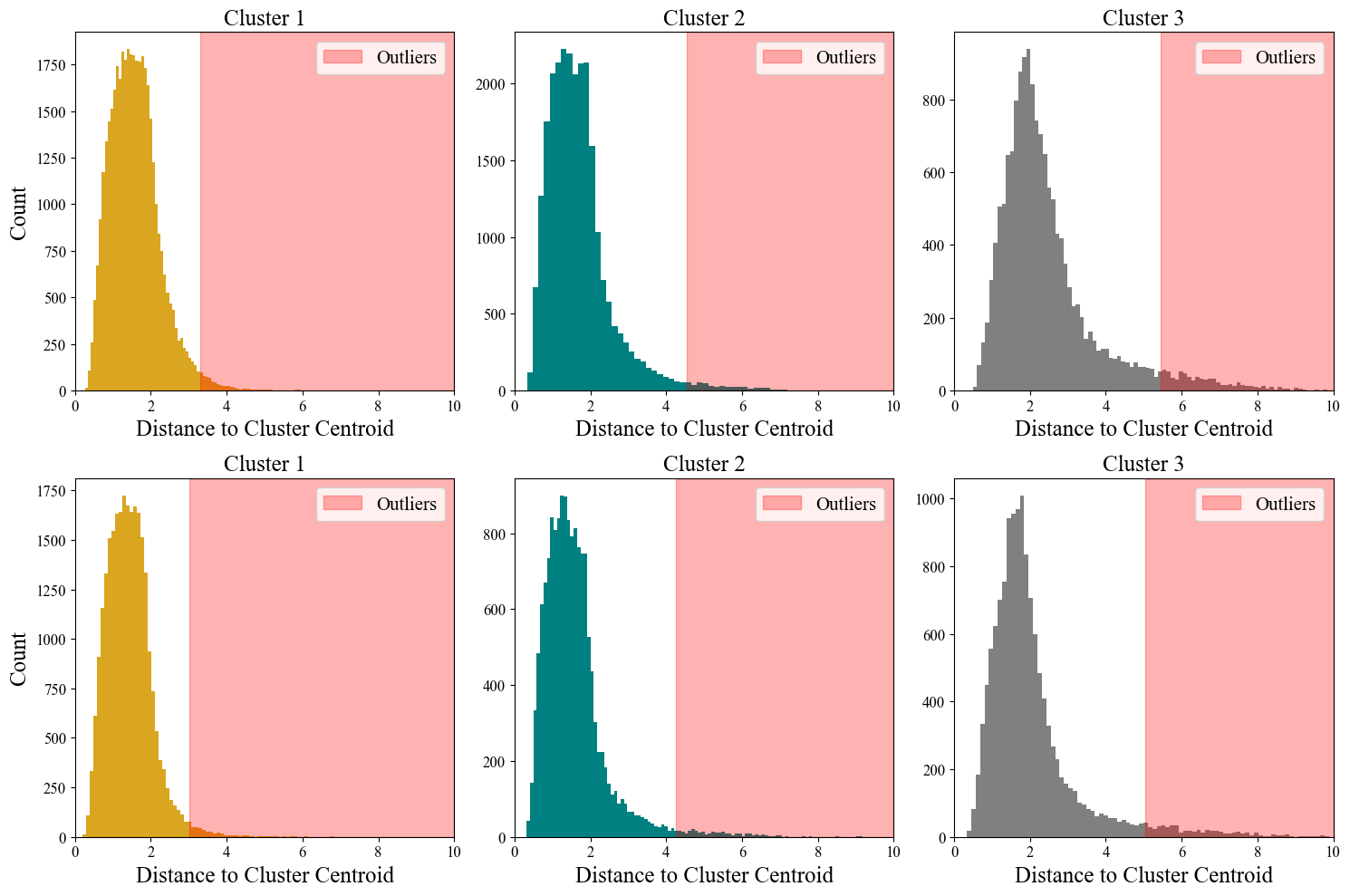}
    \caption{Histograms  show the distribution of the euclidean distance of each point from its respective cluster centroid for the {\tt\string Full} (above panel ) and {\tt\string Non-BAL Only} (below panel) datasets respectively. The red shaded region marks the respective threshold limits ($5\sigma$ for cluster 1 and 2; $4\sigma$ for cluster 3). Quasars falling within the shaded region are identified as anomalous quasars.}
    \label{fig:Distance_Histogram_Comparison}
\end{figure*}

After assigning data points to the three clusters, we computed the Euclidean distance of each point from its respective cluster centroid and analyzed the resulting distribution. Typically, spectra with coefficients similar to the majority (considered \enquote{normal} spectra) are positioned close to their cluster centroid. For clusters 1 and 2, we classified data points beyond a $5\sigma$ distance from their centroids as outliers. The $5\sigma$ threshold was chosen to ensure that the shaded region for clusters 1 and 2 in Fig.~\ref{fig:Distance_Histogram_Comparison} captures the tapering tail of the histogram. For Cluster 3, however, we applied a $4\sigma$ threshold due to its longer tail, as seen in Fig.~\ref{fig:Distance_Histogram_Comparison}. Cluster 3 (gray scatter points in the top panel of Fig.~\ref{fig:Clusters with Anomalies and mean cluster spectra}) exhibits a more diffuse and sparse spread compared to the relatively compact distributions of the other two groups, necessitating a smaller $\sigma$ cut to capture the entire tail of its histogram. We experimented with different sigma thresholds and found that a 5$\sigma$ cut for clusters 1 and 2, and a 4$\sigma$ cut for cluster 3 picks out anomalous quasars effectively as shown by the red shaded regions in Fig.~\ref{fig:Distance_Histogram_Comparison}

We obtain a total of 1994 and 1270 anomalies in the {\tt \string Full Dataset} and {\tt\string Non-BAL Only Dataset} respectively. From hereon, we refer to the quasars in these samples as \enquote{anomalous quasars}. 

\subsection{Anomaly Grouping} \label{subsec:Anomaly Grouping}

\begin{figure}[h!]
    \centering
    \includegraphics[width=\hsize]{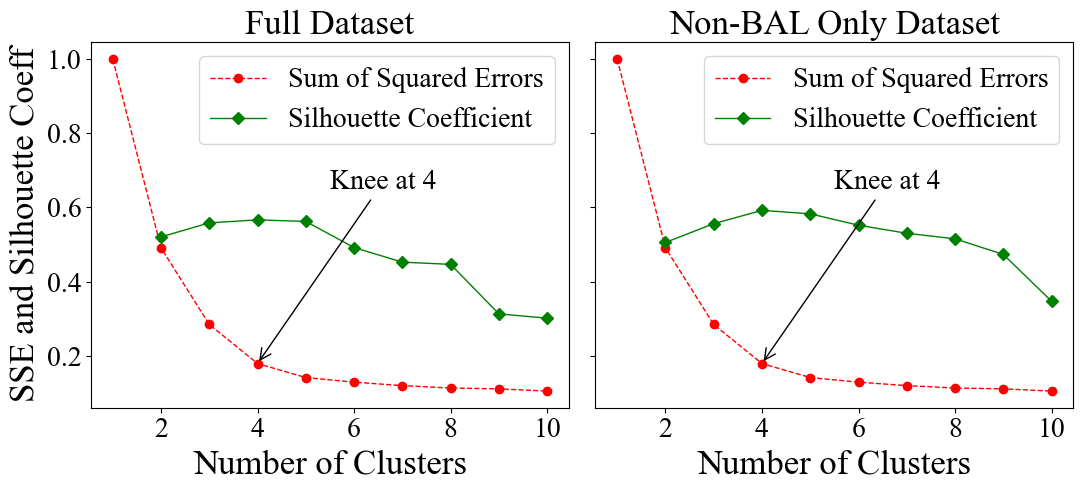}
    \caption{The figure shows the Sum of Squared Errors (SSE) and Silhouette Coefficients as a function of cluster numbers for the K-Means clustering of anomalous quasars in the {\tt\string Full}  and {\tt\string Non-BAL Only}  datasets.  The knee of the SSE curve occurs at 4 for both datasets, which coincides with the maxima of Silhouette coefficient.}
    \label{fig:Optimum Anomaly Clusters}
\end{figure}

Visual inspection of the anomalous spectra revealed recurring characteristics, such as extremely sharp and narrow \civ peaks, as well as defects like faulty spectra. To further investigate these anomalies and derive statistical insights, we re-applied K-Means clustering only on the \enquote{anomalous quasars} sample identified in \S\ref{subsec: Anomaly Detection}, to group the anomalous spectra into similar clusters, following the methodology outlined in \S \ref{subsec: PCA} and \S \ref{sec:Clustering}. For both datasets, the 20-component PCA accounted for approximately 97\% of the variance, which is sufficiently comprehensive to include nearly all significant data variance. Consequently, we retained all spectra, including those with minimal explained variance. Using the elbow method, the optimal number of clusters was determined to be 4 for the anomalies of both the datasets (see Fig.~\ref{fig:Optimum Anomaly Clusters}). Hence we grouped the anomalies of both the datasets into four groups each. 

\subsubsection{Secondary Anomaly Grouping} \label{subsubsec:Secondary Grouping}
The results showed that out of the four groups formed, two — Group 1 (G1) and Group 2 (G2) (see top panel of Fig.~\ref{fig:Anomaly_Clusters_and_Anomaly_Spectra})— were identified as \enquote{pure} groups. Here, a \enquote{pure} group refers to a population where all members exhibit the same distinct anomaly trend or feature, without further variation or subdivision within the group. In other words, every quasar in Group 1 shares the same anomaly: a relatively flat, weak continuum with a sharply peaked and narrow \civ emission line. Similarly, all quasars in Group 2 display a strongly sloped blue continuum with \civ and \siiv emission lines of nearly equal strength.  Additionally, G1 [232 members] and G2 [268 members] of the {\tt\string Full Dataset} were found to be qualitatively identical to G1 [167 members] and G2 [206 members] respectively of the {\tt\string Non-BAL Only Dataset}.

On the other hand, Groups 3 (G3) and Group 4 (G4) of the {\tt\string Full Dataset} were found to be consisting of a diverse sub-variety within the group. A typical G3 quasar from the {\tt\string Full Dataset} exhibited strong high-ionization absorption lines along with a mildly reddened continuum. In contrast, a typical G4 quasar from the same dataset showed signs of heavy reddening, with a noticeably broader \ciii\ and \mgii emission lines. The primary distinguishing feature among the sub-groups within G3 and G4 is the spectral slope, with the quasars exhibiting a range of slopes: some have a steep positive slope, others maintain a flat spectrum (indicating nearly constant flux across the wavelength range), and a few have a steeply negative slope. Furthermore, G3 and G4 from the {\tt\string Non-BAL Only Dataset}—with 336 and 207 members, respectively—appear to be subsets of their counterparts in the {\tt\string Full Dataset}, which contain 643 and 399 members respectively. 

As our primary goal is to characterize the different groups of anomalous quasars, we combined the G3 of both the {\tt\string Full Dataset} and {\tt\string Non-BAL Only Dataset}  into a composite \enquote{Group 3}. Similarly, a composite Group 4 was also created. In our analysis, we found that reddening and BAL absorption features are two prominent contributors to the observed spectral variability in quasars. Among BAL quasars, we also observed significant trends related to reddening. To better capture these variability patterns and isolate anomalies, the composite Groups were then divided into two parts each based on the {\tt\string BAL\_PROB} parameter, thereby separating the BAL anomalies and Non-BAL anomalies. We call these datasets as \enquote{Composite Group - 3/4 BAL} dataset and \enquote{Composite Group - 3/4 Non-BAL} dataset.   A third K-Means clustering, as discussed in previous sections, was performed on these BAL and Non-BAL anomalies separately, in an attempt to group the anomalous quasars. The same was done for the composite Group 4. There may be some overlap between different groups in the PCA coefficient space,  particularly for Group 3 and Group 4, which are more sparsely populated. Since K-means clustering relies solely on Euclidean distances, this can lead to certain quasars being assigned to incorrect groups during the anomaly grouping process. This overall procedure yielded four types of BAL anomalies as described in \S\ref{discussion:bal anomalies} and three types of Non-BAL Anomalies, later found to be Reddened Anomalies as discussed in \S\ref{result:red anomalies}.

Fig.~\ref{fig:flowchart} presents a flowchart outlining the steps followed by our algorithm, starting from quasar sample selection and progressing through to the final identified anomaly groups.

\section{Results} \label{sec:Results}

It is important to note that, hereafter, the three classifications resulting from the initial K-Means clustering on the entire dataset will be referred to as \textbf{Clusters}, while the classifications of anomalous spectra from the subsequent K-Means clustering will be referred to as \textbf{Groups}.

\subsection{Clusters}\label{subsec: Results_Clusters}
\begin{figure*}[h!]
    \centering
    \begin{tabular}{c}
    \includegraphics[width=\hsize]{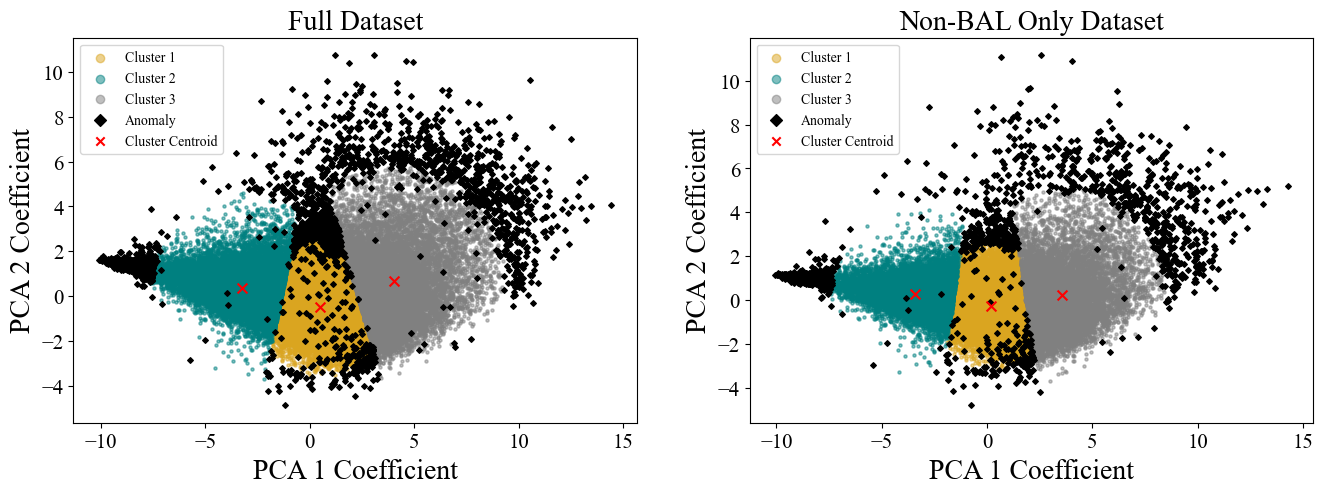} \\
    \includegraphics[width=\hsize]{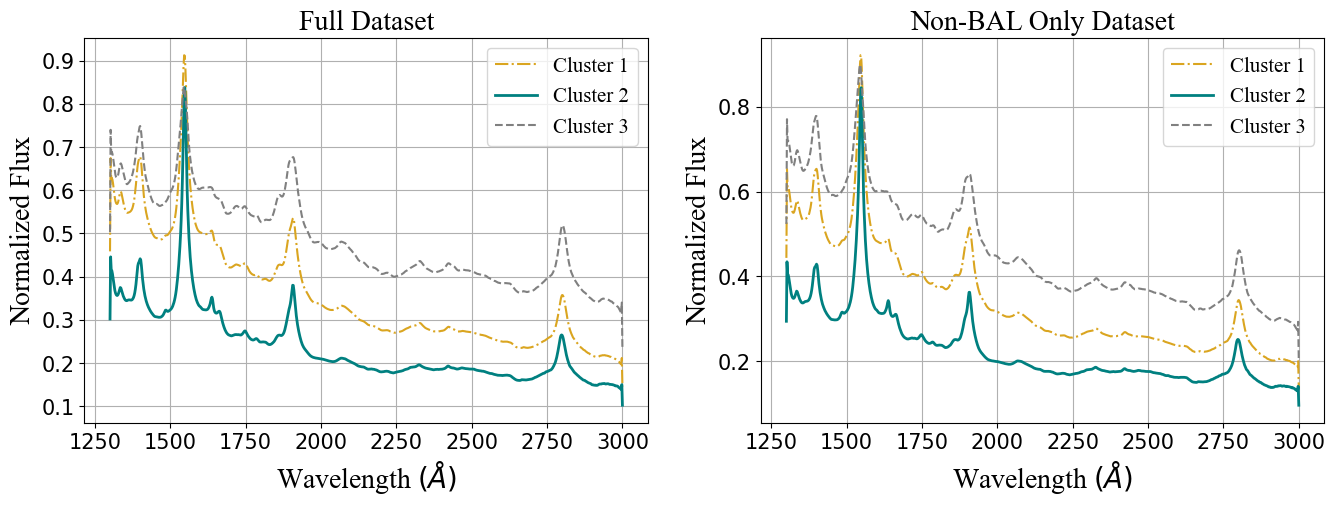} \\
    \end{tabular}
    \caption{Top Panel : The figure presents a 2D projection (PCA 1 versus PCA 2 coefficients) of the {\tt\string Full} (left)  and {\tt\string Non-BAL Only} (right) datasets. Each dataset is divided into three clusters (cluster 1: yellow, cluster 2: teal, cluster 3: gray) within the 20-dimensional PCA hyperspace, using K-Means clustering. Quasars classified as anomalous after applying a $5\sigma$ threshold to clusters 1 and 2, and a $4\sigma$ threshold to cluster 3, are shown as black scatter points overlayed on top of the cluster members. The cluster centroids are marked by red crosses. Bottom Panel: Mean composite spectrum for each cluster of the {\tt\string Full} (left)  and {\tt\string Non-BAL Only} (right) datasets. The color of each spectrum corresponds to the color of the cluster as shown in the top panel.}
    \label{fig:Clusters with Anomalies and mean cluster spectra}
\end{figure*}

As shown in Fig.~\ref{fig:PCA_Explained_Variance}, the first three principal components account for nearly all the explained variance. Therefore, the K-Means clusters are visualized by plotting the first two PCA coefficients. The top panel of Fig.~\ref{fig:Clusters with Anomalies and mean cluster spectra} shows the three clusters identified by the initial clustering algorithm, visualized in the PCA 1 versus PCA 2 coefficient space for both the {\tt\string Full Dataset} (left) and the {\tt\string Non-BAL Only Dataset} (right).  Outliers identified, as described in \S \ref{subsec: Anomaly Detection}, are highlighted in black, in the same panel. Clusters 1 and 2 appear relatively compact with distinct boundaries along both axes, while Cluster 3 shows a broader spread, particularly along the second PCA eigenvector.

The number of anomalies present in each cluster is given in Table \ref{table:Anomalies per cluster}:
\begin{table}[h!]
\begin{tabular}{|c | c | c | c|} 
 \hline
 Dataset & Cluster 1 & Cluster 2 & Cluster 3 \\ [0.5ex] 
 \hline
 {\tt\string Full} & 702 & 494 & 283 \\ 
 {\tt\string Non-BAL} & 411 & 296 & 271 \\
 \hline
\end{tabular}
\caption{Number of anomalies in each cluster for both datasets}
\label{table:Anomalies per cluster}
\end{table}
Some anomalies may appear to be \enquote{within} the clusters in the  2D visualization; however, this is a result of projecting the 20-dimensional hyperspace onto two dimensions. Points assigned to two distinct clusters that appear to overlap in the 2D projection may actually be distant from each other in the higher-dimensional space, thereby qualifying as outliers in the full dimensional space. \\
Cluster 1 contains 4138 BAL quasars, and Cluster 2 has 2779 BAL quasars, constituting about 30\% of the total quasars in each cluster. Cluster 3 on the other hand has 1861 BAL quasars which accounts for nearly half $(48.6\%)$ of the quasars in that cluster.\\
To better understand the basis of clustering, we created mean or composite spectra \citep[e.g.][]{2001AJ....122..549V} for each cluster, which highlight the average spectral properties of the members in each cluster, as shown in the bottom panel of Fig.~\ref{fig:Clusters with Anomalies and mean cluster spectra}. This plot provides a visual representation of why the quasars were classified into three distinct categories.

Clearly, the three clusters have  different spectral slopes. Cluster 3 (gray dashed line in the bottom panel of Fig.~\ref{fig:Clusters with Anomalies and mean cluster spectra}) has the steepest continuum, while Cluster 2 (teal solid line) exhibits the flattest mean spectrum. Notably, the \civ emission line in Cluster 2 is significantly stronger and narrower than in the other clusters. Conversely, the iron emission between 2250 and 2500 \AA\, is weakest in Cluster 2 but strongest in Cluster 3, where the characteristic double-hump feature is clearly visible. Additionally, the blueward wing of the \ciii emission line appears significantly stronger in Clusters 2 and 3,  while in Cluster 1 it shows a more symmetric profile. The clusters also differ in average flux, with Cluster 3 having the highest average flux and Cluster 2 the lowest.

When comparing  the  {\tt\string Full Dataset} and {\tt\string Non-BAL Only Dataset}, the composite spectrum of cluster 2 is found to be nearly identical in both. Since the {\tt\string Full Dataset} includes BAL quasars, whereas the {\tt\string Non-BAL Only Dataset} does not, the similarity between the spectra of cluster 2 in both datasets suggests that BAL features are not prominent within this cluster. However, it's important to note that cluster 2 still contains around 30\% BAL quasars, which is a significant fraction.  We rather think that the extremely sharp peak of the \civ emission line in these quasars, when max-normalized, reduces the prominence of BAL features. The strong \civ emission may dominate the mean spectrum, effectively masking the impact of the BAL absorption lines and leading to their reduced prominence in the overall spectrum. On the other hand, cluster 1 and 3 of the {\tt\string Non-BAL Only Dataset} show higher flux emission towards the high energy end of the spectrum (between 1250 -- 1500 \AA) as compared to the cluster 1 and 3 respectively of {\tt\string Full Dataset}. This indicates that the BAL quasars reduce the flux at high energy end (1250 -- 1500 \AA) which reduces the mean flux blueward of the \civ emission line. 

\subsection{Groups}
\begin{figure*}[h!]
    \centering
    \begin{tabular}{c}
    \includegraphics[width=\hsize]{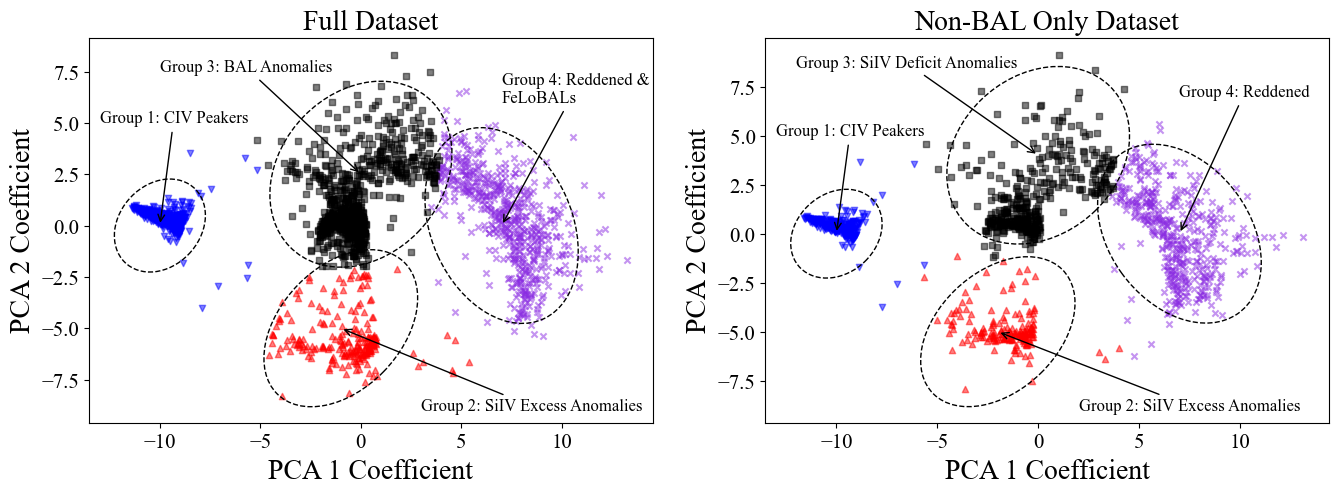}\\
    \includegraphics[width=\hsize]{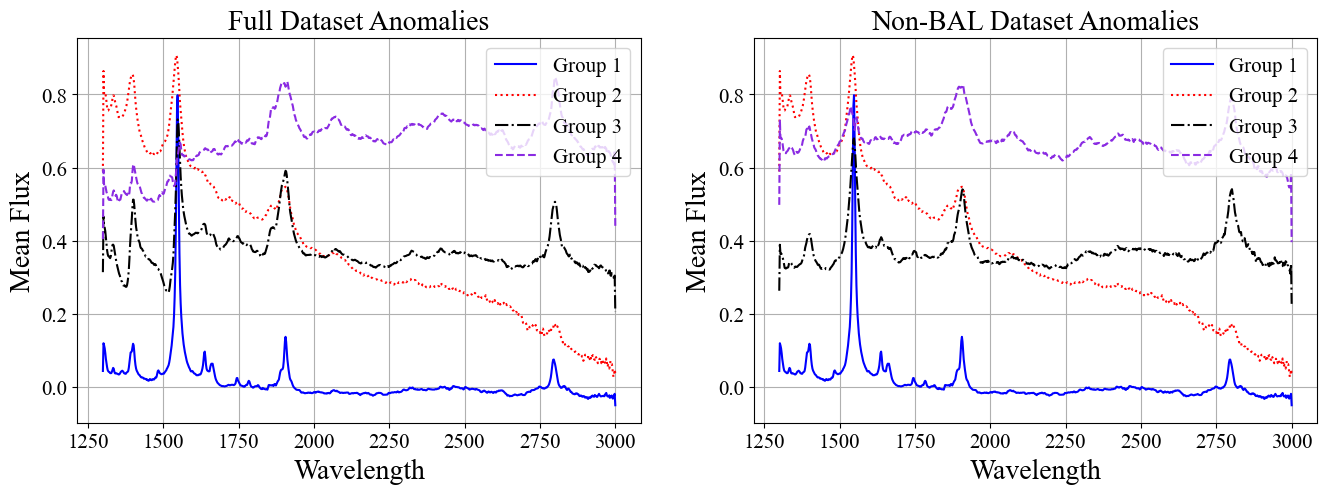}
    \end{tabular}
    \caption{Top Panel : A 2D projection (PCA 1 versus PCA 2 coefficients) of the anomalies of the two datasets as divided into four groups each (group 1: blue, group 2: red, group 3: black, group 4: violet), in the 20 dimensional PCA hyperspace, by the second K-Means clustering applied only on the anomalous quasars. The groups as characterised by further analysis (see \S\ref{sec:Discussion}) are circled and labelled to reflect the nature of their anomaly. Bottom Panel : Mean composite spectrum of each anomaly group for both the datasets. The colors of the spectra correspond to the  color of the corresponding group in the top panel.}
    \label{fig:Anomaly_Clusters_and_Anomaly_Spectra}
\end{figure*}

As detailed in \S \ref{subsec:Anomaly Grouping}, K-Means clustering was reapplied to the detected anomalies (points marked in black in the top panel of Fig.~\ref{fig:Clusters with Anomalies and mean cluster spectra}), resulting in the formation of 4  groups for the {\tt\string Full} and {\tt\string Non-BAL Only} datasets. The visualization of these anomaly groups (excluding cluster members) using the first two PCA coefficients is presented in the top panel of Fig.~\ref{fig:Anomaly_Clusters_and_Anomaly_Spectra}, while the bottom panel displays corresponding composite spectra for each group. The colors of each spectrum in the bottom panel of Fig.~\ref{fig:Anomaly_Clusters_and_Anomaly_Spectra} are chosen to match the color scheme in the top panel. Additionally, we compared individual anomalous spectra against cluster and group composites, to ensure that spectra in a group are similar. 

\begin{table}[h!]
\centering
\begin{tabular}{|c | c | c | c| c|} 
 \hline
 Dataset & Group 1 & Group 2 & Group 3 & Group 4\\ [0.5ex] 
 \hline
 {\tt\string Full} & 232 & 268 & 643 & 399 \\ 
 {\tt\string Non-BAL} & 167 & 206 & 336 & 207\\
 \hline
\end{tabular}
\caption{Number of anomalies in each group for both datasets}
\label{table:Anomalies per group}
\end{table}
 The exact number of anomalies in each group is given in Table \ref{table:Anomalies per group}. Based on these composite group spectra, the four groups created by the second K-Means clustering can be seen as:

\subsubsection{Group 1 - Full/Non-BAL Only Datasets}
This group features quasars with an extremely sharp and narrow \civ emission line, consistently identified across both datasets, with identical group members in each. They occupy the same location in the PCA 1 versus PCA 2 plot for both datasets, specifically in the leftmost region of the distribution, marked by blue dots in the top panel of  Fig.~\ref{fig:Anomaly_Clusters_and_Anomaly_Spectra}. Hereafter, this group will be referred to as \civ Peakers.  Notably, this group also contains  contaminant quasar spectra ($\sim350)$ characterized by a sharp, narrow artifact peak —likely caused by cosmic ray encounters— which mimics the properties of the narrow \civ peak, leading to coincident grouping. These cosmic ray anomalies are easily removed by placing an upper cut on the \civ equivalent width (EW) and flux ($\lambda1549$\AA) since \civ Peakers are known to have relatively high \civ EW and flux (see \S\ref{subsec:CIV Peakers}) conveniently separating them from the cosmic ray anomalies. There are a total of 65 \civ Peaker anomalies that account for 0.08\% of the total selected quasars and represents approximately 3.4\% of the total detected anomalies. 

\subsubsection{Group 2 - Full/Non-BAL Only Datasets}
This group is identified in both datasets and is depicted by the red scatter in the top panel of Fig.~\ref{fig:Anomaly_Clusters_and_Anomaly_Spectra}, located in the bottom-most region of the PCA distribution. The low PCA 2 coefficients indicate a lack of reddening and a relatively blue spectrum. Consequently, these quasars are characterized by a strongly downward-sloping continuum with equally strong \civ and \siiv emission lines.  Hereafter, we call these quasars as Excess \siiv emitters.
There are 227 Excess \siiv Emitters, representing 0.27\% of the total quasars in our sample and approximately 11.3\% of the total anomalies.  This group also features the identification of 13 blazars, which lack any emission line but have a similar steep downward sloping spectrum as a typical \siiv Excess anomaly (refer to panel (c) in Fig.~\ref{fig:all anomaly spectra}) for the mean spectrum of the 13 blazars). These blazars are present in the lower extreme of the \siiv excess anomalies spread in the PCA eigenspace in the top panel of Fig.~\ref{fig:Anomaly_Clusters_and_Anomaly_Spectra}. This group in the {\tt \string Full Dataset} also contains rare BAL quasars with a bluer continuum. These quasars were manually identified and reassigned to the appropriate sub-group within Group 3. These are discussed in  detail in \S\ref{blue bals}.

\subsubsection{Composite Group 3 - BAL QSOs} \label{result:BALAnomalies}
\begin{figure*}[h!]
    \centering
    \begin{tabular}{c c}
    BAL Anomalies & Reddened Anomalies\\
    \includegraphics[width=0.475\hsize]{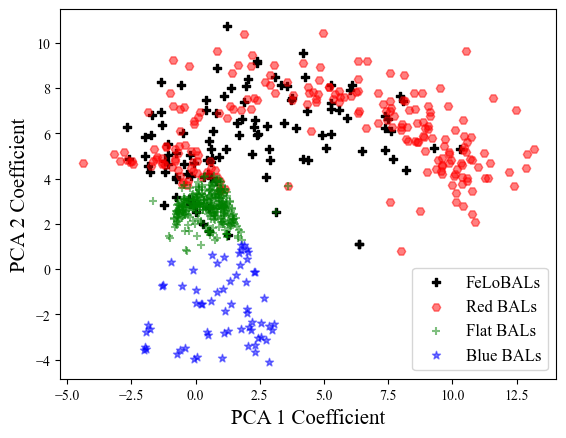} &
    \includegraphics[width=0.475\hsize]{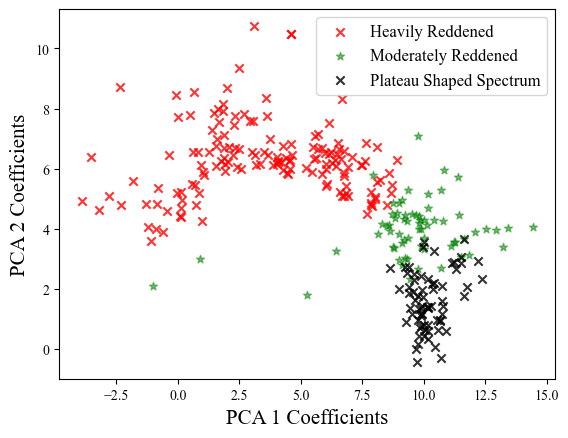}
    \end{tabular}
    \caption{Top Panel : 2D projection of BAL anomalies (using PCA 1 versus PCA 2 coefficients) as grouped into 4 types by the secondary K-Means clustering as discussed in \S\ref{subsubsec:Secondary Grouping}. The Reddened and FeLoBALs are placed on the upper region with higher PCA 2 coefficients pertaining to largely amount of reddening in their spectra. Flat BALs, usually with a flat spectrum are placed in the middle region. Blue BALs comprising of primarily High Ionization Broad Absorption Line Quasars (HiBALs) are placed in the lower region of the distribution with minimum PCA 2 coefficients. Bottom Panel : 2D projection of reddened anomalies (using PCA 1 versus PCA 2 coefficients) as grouped into 3 sub-types by the secondary K-Means clustering as discussed in \S\ref{subsubsec:Secondary Grouping}. The heavily reddened quasars, characterized by steeply sloped spectra, are positioned in the upper region with higher PCA 2 coefficients. Moderately reddened quasars, exhibiting a flat continuum, occupy the middle region. Quasars with plateau-shaped spectra are located in the lower region of the distribution, with the lowest PCA 2 coefficients, due to their overall negative slope.}
    \label{fig:BAL and Red Spread}
\end{figure*}

BAL QSOs are spread throughout the PCA eigenspace with maximum density in the Cluster 1 (see top panel of Fig.~\ref{fig:Clusters with Anomalies and mean cluster spectra}). However, a significant number of BAL quasar spectra are also classified as anomalous. In fact, this is the largest anomaly group, comprising 672 objects, which account for 38\% of all detected anomalies and 0.8\% of the total dataset. These anomalies are present only in the {\tt\string Full Dataset} and are divided into four subgroups through secondary grouping: (1) Blue BALs: Quasars with strong high-ionization absorption lines and a steep downward-sloping (blue) continuum. This is notable since typical BAL quasars usually show significant reddening,
(2) Flat BALs:  BAL quasars with a relatively flat continuum. About 80\% of Flat BALs are low-ionization BALs (LoBALs), which are BAL quasars that feature absorption lines from low-ionization species like \mgii and \aliii, along with high-ionization lines,
(3) Red BALs: BAL quasars with an upward-sloping (red) spectrum, indicating significantly more reddening than typical BAL quasar spectra, and (4) FeLoBALs: A subset of LoBALs with absorption lines from iron transitions (excited/ground state), known as FeLoBALs (the exact number of quasars in each of these anomaly classes is given in Table \ref{table:bal anomalies per group}).
In the PCA eigenspace, these four subgroups are primarily separated by the degree of spectral reddening. This separation is visible in the left panel of Fig.~\ref{fig:BAL and Red Spread}, which shows the spatial distribution of all BAL anomalies within the PCA space.

\subsubsection{Composite Group 3 - Non BAL QSOs Only}
This group appears as a distinct, pure group within the {\tt\string Non-BAL Only} dataset anomalies  (black scatter points in the top panel of Fig.~\ref{fig:Anomaly_Clusters_and_Anomaly_Spectra}). They are located in the topmost region of the PCA 1 versus PCA 2 plot, indicating significant reddening in their continuum, which results in a relatively flat spectral slope.  The members of this group are characterised by a disproportionately low emission in the \siiv emission line. Hereafter, we call these quasars as \siiv deficient anomalies.   There are 328 \siiv deficient anomalies which account for 0.4\% of the total quasars in our sample and 16.4\% of all anomalies. \siiv deficient anomalies appear as a part of the BAL anomalies in the {\tt\string Non-BAL Only} dataset. 

Although the {\tt\string Non-BAL Only Dataset} was expected to be free of any BAL quasars, this group contains 18 BAL quasars that were missed by the BAL detection pipeline of \citet{Guo_2019} and incorrectly assigned a zero {\tt\string BAL\_PROB} value. These \enquote{leaked BALs} have a spectral shape similar to that of the members of this group, which led to their inclusion.  These previously undetected BAL quasars were manually picked out by visual inspection and subsequently flagged for redaction and proper classification.

\begin{table}[h!]
\centering
\begin{tabular}{|c |  c  c  c c|} 
 \hline
 Type & Blue BALs & Flat BALs & Red BALs & FeLoBALs\\ [0.5ex] 
 \hline
 {\tt\string Count} & 64 & 306 & 213 & 109 \\ 
 \hline
\end{tabular}
\caption{Number of members in each BAL anomaly sub-group}
\label{table:bal anomalies per group}
\end{table}

\subsubsection{Composite Group 4 - BAL QSOs} \label{result:red anomalies}
This anomaly group is placed at the upper right section of the PCA 1 versus PCA 2 plot (see top panel of Fig.~\ref{fig:Anomaly_Clusters_and_Anomaly_Spectra}), indicating maximum values of PCA 1 and PCA 2 coefficients. These large values correspond to the strong reddening observed in its members. Upon secondary grouping as described in \S\ref{subsubsec:Secondary Grouping}, this group was divided into two parts, i.e. Red BALs and FeLoBALs. The FeLoBAL population found here exhibits a greater amount of spectral reddening as compared to the FeLoBALs found in Composite Group 3 - BALs (\S\ref{result:BALAnomalies}). This distinction is primarily based on the amount of reddening in the quasar and hence the overall spectral shape (slope). Additionally, the strength and width of the absorption lines in the quasar spectra also serves as an important distinguishing property amongst the subgroups. The FeLoBALs make up nearly 50\% of this group while the other half consists of Red BALs. A detailed discussion about these subgroups is given in \S\ref{reddened anomalies}.

\subsection{Composite Group 4 - Non-BAL Only}
\begin{figure*}[h!]
    \centering
    \includegraphics[width=\hsize]{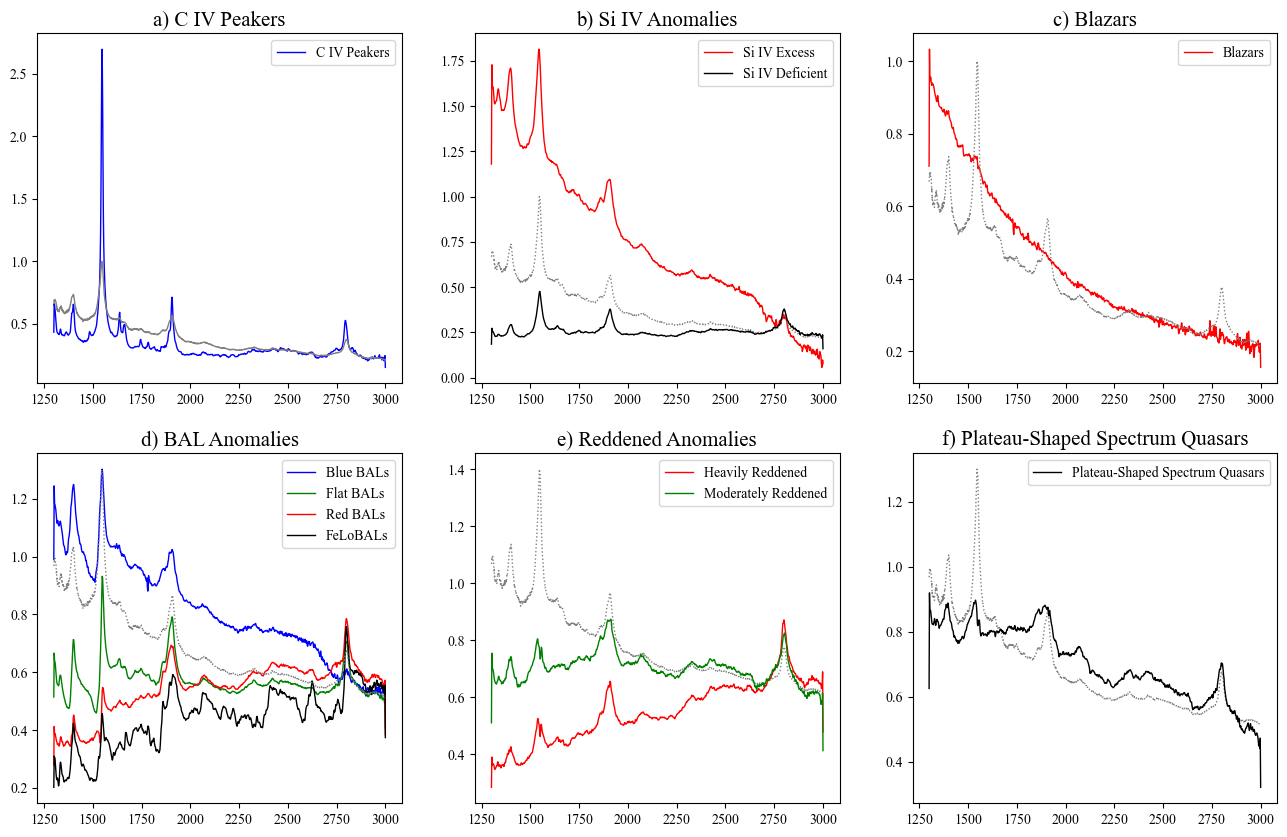}
    \caption{The figure shows the composite spectra for all anomaly categories identified in this project. In each plot, the gray dotted spectrum represents the \citet{2001AJ....122..549V} composite. All mean spectra are aligned at the \mgii emission peak to ensure consistent comparison across categories.}
    \label{fig:all anomaly spectra}
\end{figure*}

Similar to the group 4 of the {\tt\string Full Dataset}, this group is also present at the top right part of the PCA 1 versus PCA 2. Upon secondary K-Means clustering (\S\ref{subsubsec:Secondary Grouping}) this group was divided into three parts i.e. Heavily Reddened, Moderately Reddened and Plateau-Shaped spectrum quasars. While the heavily reddened quasars have a steep positively sloping spectrum, the moderately reddened quasars have a relatively flat continuum. The third subgroup here, is a new type of quasar population, which we refer to as \enquote{Plateau-shaped spectrum Quasars}. Plateau-shaped spectrum quasars exhibit a spectral break around 2000 \AA. Blueward of this break, their spectra display a flatter continuum, while redward of the break, the spectrum becomes steeper and bluer (see panel (f) in Fig.~\ref{fig:all anomaly spectra}). These Plateau-shaped spectrum quasars are present in the lowermost section of the group's PCA spread (see black scatter points in the right panel of Fig.~\ref{fig:BAL and Red Spread}) due to their relatively bluer continua. Heavily reddened quasars constitute 60 percent of the total population of this group while the rest 40 percent is equally made of moderately reddened quasars and Plateau-shaped spectrum quasars. 

\section{Discussion} \label{sec:Discussion}

We identify ten distinct groups of spectroscopic anomalies in the SDSS DR16 Quasar catalog by applying hierarchical unsupervised K-Means clustering to the spectral PCA decompositions. BAL quasars are well known for having spectra that differ from normal quasars due to the presence of strong absorption features. To prevent BAL quasars from being identified as major anomalous groups, we analyzed two datasets: one containing BAL quasars {\tt\string Full Dataset} and another devoid of BAL quasars i.e the {\tt\string Non-BAL Only Dataset}. This approach classified 81,814 quasars into three clusters, ensuring that quasars within each cluster have similar spectral characteristics, regardless of their luminosities.

The composite spectra of these clusters exhibit several properties that contribute to their unique separation. Cluster 2 (see top panel of Fig.~\ref{fig:Clusters with Anomalies and mean cluster spectra}) in both the datasets exhibits nearly identical spectrum as the striking, extremely strong \civ peak dilutes the impact of BAL signature absorption on the spectrum. The anomaly clusters in both datasets show a similar distribution and spread across all groups, with the \control\ containing a larger number of data points. The inclusion of BAL quasars in the \control\ results in more diffuse and broader spectral clusters, likely due to the strong absorption lines characteristic of BAL quasars. In contrast, their removal in the \nobal\ produces more distinct and narrower spectral clusters, suggesting that BAL features contribute significantly to spectral variability, particularly at the bluer end of the UV-optical spectrum.

This observation validates our approach of analyzing two datasets: one with all spectra and another with BAL quasars removed. This strategy proves effective, as the Reddened and \siiv Deficient quasar anomaly groups in the \nobal\ appear as minor contaminants within the  FeLoBAL and BAL Anomaly groups in the \control, respectively. By removing BAL quasars from the \control\, to create the \nobal dataset, we were able to detect other anomalous quasars that might otherwise be overshadowed.

In the following section, we analyze the distribution of equivalent widths, line ratios, and Full Width at Half Maximum (FWHM) \citep[from][]{Wu_2022} for prominent lines across all anomaly groups to gain deeper insights into the underlying physics of each group. Inferences are drawn from the analyzed distributions and attributed to specific physical properties, as supported by findings from previous studies. A detailed characterization of each anomaly group, including multi-wavelength and multi-epoch analysis of individual objects, is outside the scope of this paper and will be explored in future studies. 

\subsection{{C\,\,{\sc iv}~}Peakers} \label{subsec:CIV Peakers}

\begin{figure*}[h!]
    \centering
    \includegraphics[width=\hsize]{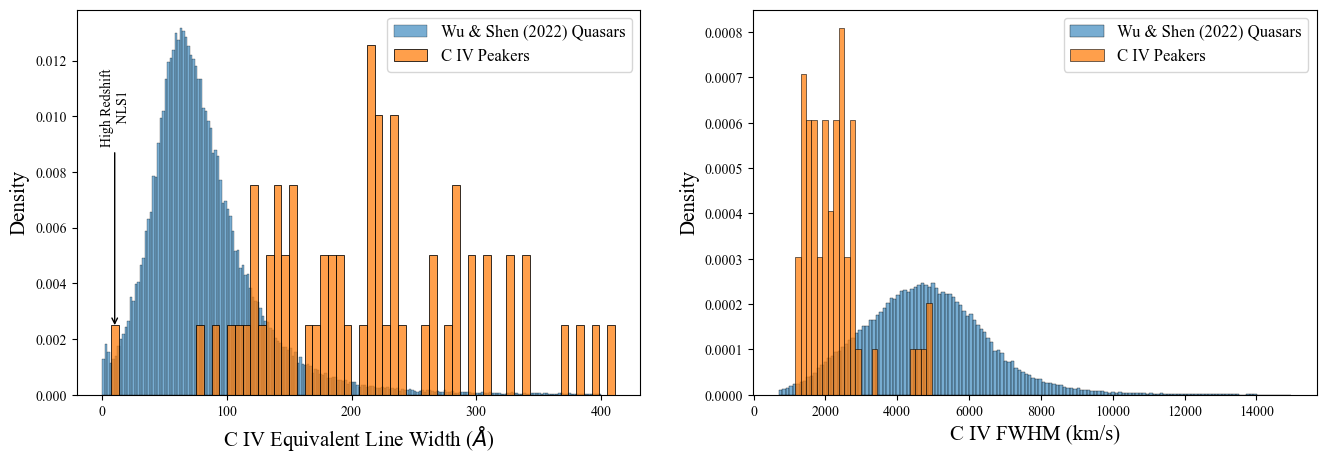}
    \caption{The figure shows the distribution of \civ equivalent width (EW) and full width at half maximum (FWHM) for all quasars in the \citet{Wu_2022} catalog (blue) compared to the \civ Peakers (orange). The \civ EW for the \civ Peakers is concentrated at the higher end of the distribution, while their FWHM is generally lower, indicating a strong yet narrow \civ emission line.}
    \label{fig:CIV Peaker Properties}
\end{figure*}

Members of this group are characterized by an exceptionally strong and narrow \civ emission line, accompanied by a weak, flat continuum (see panel (a) in Fig.~\ref{fig:all anomaly spectra}). Fig.~\ref{fig:CIV Peaker Properties} shows the comparison of the \civ EW (left;(a)) and \civ FWHM (right;(b)) of the \civ Peakers to that of all the quasars in the \citet{Wu_2022} catalog. For this group the median \civ FWHM is $2121.46\pm175.55\,$\kms\,which is 0.53 times smaller than the median \civ FWHM of all the quasars  in the \citet{Wu_2022} catalog. Additionally, the median FWHMs of all other emission lines ({Si\,{\sc iv}}, {He\,{\sc ii}}, {C\,{\sc iii}}, {Mg\,{\sc ii}}) are also lower, ranging from 0.4 to 0.7 times their respective medians  in the\citet{Wu_2022} catalog. 

In terms of equivalent width, the median \civ EW is $151.33\pm3.24\,$\AA\, which is $\sim$ 3.3 times the median \civ EW of all the quasars in \citet{Wu_2022}. 
Except for the leftmost point in the Fig.~\ref{fig:CIV Peaker Properties}b, which was identified as a Narrow Line Seyfert 1 (NLS1) galaxy as documented by \citet{rakshit2021high}, the majority of the \civ Peakers occupy the high EW end of the distribution. Despite the extremely high EWs, the \civ FWHM for these quasars remains on the lower end of the distribution, indicating that the extreme EW is driven by their exceptional line strength. However, five samples display relatively high FWHM alongside sharply peaked \civ emission (see Fig.~\ref{fig:CIV Peaker Properties}b). 

Additionally, along with \civ, the median EW of the \heii line is about 2.31 times stronger than the median \heii EW of all quasars in \citet{Wu_2022}. However, the median EW values for {Si{\sc iv}}, {C{\sc iii}}, and {Mg{\sc ii}} align closely with the median values observed in \citet{Wu_2022}. The flux ratio, defined as the ratio of integrated fluxes between two lines, and the EW ratio of any line relative to \civ for this group are positioned at the extremely low end of the distribution compared to all quasars in \citet{Wu_2022}, due to the exceptionally high strength of the \civ line.
This suggests that the anomalous behavior is primarily driven by the extraordinary strength and EW of the \civ emission line.  The \civ emission line stands out as the most prominent feature within the chosen wavelength window in our analysis for nearly all quasars (except for a few BALs), underscoring the algorithm's effectiveness in identifying this strong line characteristic.

The \civ emission line, located at the higher end of the UV spectrum, predominantly originates from the accretion disk near the central black hole \citep{Peterson_1997}. It arises from the $^2P_{3/2,1/2}\rightarrow\,^2S_{1/2}$ (C\textsuperscript{+++} ground state) transition. As a major coolant for gas at high temperatures, well above 10\textsuperscript{4}K \citep{ferland1996high}, the \civ emission is significantly enhanced. \civ $\lambda$1550\AA\ has an ionization potential of 64 eV, while \heii $\lambda$1640\AA\ has an ionization potential of 54 eV. Strong high-ionization lines like \civ  and \heii  are typically indicative of a high ionization parameter, $\Gamma \geq 10^{-2}$ or harder continua \citep[see][]{marziani1996comparative}.

A comprehensive explanation for the extremely high \civ flux and equivalent line width  is provided by \citet{fu2022nature}(hereafter F22), who conducted UV-Optical and X-ray analyses of eight quasars with very high \civ EW ($>$150 \AA). Notably, two of these quasars are present in our group. In the F22 sample, large HeII EWs are also observed, indicating that the high-ionization broad-line region (BLR) is receiving a significant number of ionizing photons, consistent with a hard ionizing continuum. 
F22 discusses the possibility that sources with large \civ EWs may represent the \enquote{opposite extreme} to Weak Line Quasars (WLQs), which typically exhibit small \civ EWs, large \civ blueshifts, and weak X-ray emission. WLQs are often explained as quasars with high Eddington ratios, resulting in a geometrically and optically thick inner accretion disk that drives outflows. The thick inner disk together with the outflows can prevent ionizing EUV/X-ray photons from reaching the high-ionization broad emission line region and, in some cases, block the line of sight to the central X-ray-emitting region \citep[e.g.][]{ni2018connecting,ni2022sensitive}. In contrast, quasars with large \civ EWs may have relatively low Eddington ratios and minimal intrinsic absorption. 
Finally, F22 concluded that the enhancement in \civ and \heii is best explained by the combined effects of a hard ionizing continuum and sub-solar metallicity. An enhanced high-ionization recombination line like \heii suggests a harder ionizing continuum. In combination with this harder continuum, lower metallicity results in \civ becoming the dominant coolant. The strength of resonance lines depends on both ionic abundance and ionizing conditions. In a lower metallicity environment, other lines are not significantly enhanced due to reduced ionic abundance, whereas \civ remains prominent as the primary coolant under these conditions.
\begin{figure*}[h!]
    \centering
    \begin{tabular}{c c}
    \siiv Excess Anomalies & \siiv Deficient Anomalies\\
    \includegraphics[width=0.475\hsize]{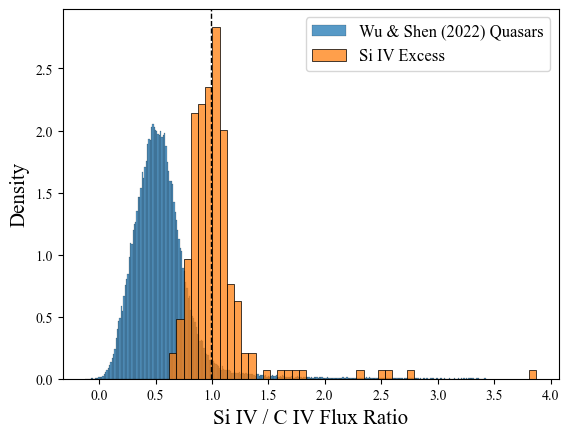} &
    \includegraphics[width=0.475\hsize]{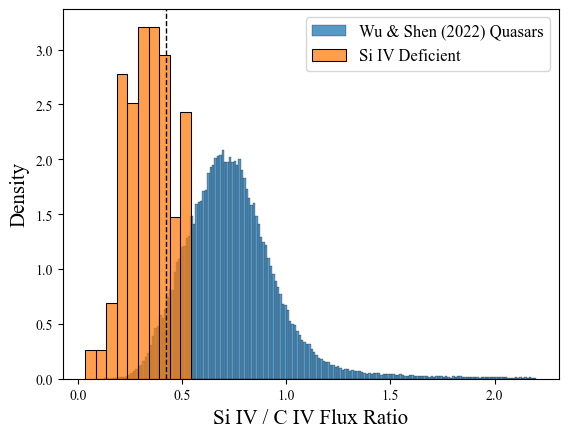}
    \end{tabular}
    \caption{Top Panel : The figure shows the distribution of the ratio of \siiv to \civ emission line flux for the Excess \siiv emitters group (orange) compared to all quasars in the \citet{Wu_2022} catalog (blue). The \siiv to \civ ratio centered around 1 signifies an enhanced \siiv emission, where the \siiv emission line is as strong as the \civ line. Bottom Panel : The figure shows the distribution of the ratio of \siiv to \civ emission line flux for the \siiv Deficient group (orange) compared to all quasars in the \citet{Wu_2022} catalog (blue).}
    \label{fig:SiIV anomaly properties}
\end{figure*}

\subsection{Excess {Si\,\,{\sc iv}} Emitters} \label{subsec: Excess SiIV Emitters}

These quasars are characterized by the presence of \siiv and \civ emission lines of nearly equal strength, accompanied by a strongly negative-sloped bluer continuum (see panel (b) in Fig.~\ref{fig:all anomaly spectra}). A total of 227 such objects were identified in our study. The median \siiv EW of this group is $9.19\pm0.46$\AA\, which is nearly equal to the median \siiv EW of all quasars in the \citet{Wu_2022} catalog. In contrast, the median EWs of other spectral lines in this group are reduced by approximately 40\% to 50\% compared to their respective median values in the \citet{Wu_2022} catalog. For this group, the EW ratio of \siiv to other lines—such as \civ, \heii, \ciii, and \mgii—is approximately twice that of the corresponding EW ratios in the \citet{Wu_2022} catalog. This suggests that the overall anomalous nature of these quasars is primarily due to the heightened \siiv emission relative to other lines.

The left panel of Fig.~\ref{fig:SiIV anomaly properties} shows the distribution of the flux ratio of \siiv to \civ emission lines for the excess \siiv emitter group compared to the distribution in the \citet{Wu_2022} catalog.

The median emission flux ratio of \siiv to \civ in the \citet{Wu_2022} catalog is approximately 0.72. However, for this group, the ratio increases to about 0.99, as indicated by the peak centered around 1 in the left panel of Fig.~\ref{fig:SiIV anomaly properties}. This shift is attributed to a combination of factors: the \siiv emission flux is 4.2 times stronger, while the \civ flux is only half as strong. As a result, the \siiv and \civ emission line peaks reach nearly the same height, and in many cases, the \siiv peak even surpasses the \civ peak in this group's spectra.\\

We note that the \heii emission line in this group is significantly broader, with a FWHM nearly three times the median \heii FWHM of all quasars in the \citet{Wu_2022} catalog. In contrast, the FWHM distributions for other lines ({C\,{\sc iv}}, {Si\,{\sc iv}}, {C\,{\sc iii}}, {Mg\,{\sc ii}}) closely match those of the quasars in the \citet{Wu_2022} catalog.

\citet{nagao2006evolution} demonstrated that the flux ratio of [\siiv+ O \textsc{iv}] to \civ is a key diagnostic for probing the physical conditions and chemical composition of gas in quasars, particularly within the BLR, as the relative role of \civ as a coolant diminishes with increasing BLR metallicity. Our analysis reveals an \siiv/\civ flux ratio of approximately 1.4 (compared to an average of ~0.5 for the overall dataset).
According to \citet{hamann2002metallicities}, such a high ratio is indicative of super-solar BLR gas metallicities. This elevated metallicity reflects the presence of heavy elements produced in star-forming regions and incorporated into the quasar’s environment, a process often referred to as \enquote{chemical enrichment}. Consequently, the elevated {Si\,{\sc iv}}/{C\,{\sc iv}} ratio in this group suggests super-solar BLR gas metallicities, possibly due to unusual stellar activity near the galactic nuclei, which could explain their anomalous nature.

\subsection{{Si\,\,{\sc iv}} Deficient Anomalies}
This group can be considered the counterpart to the \siiv Excess group discussed earlier. In these quasars, we observe reduced \siiv emission compared to \citet{Wu_2022} quasars, while the \civ emission between the two samples remains comparable.
Their spectra typically exhibit a characteristic flatter continuum. Some members ($\sim$18\%) also display notably enhanced iron emissions ($\sim44-53\%$) between 2250 and 2700 \AA.
The ratio of \siiv to \civ flux for this group is 0.23, compared to 0.75 for the entire quasars in . This ratio, which serves as a metallicity indicator for the BLR, points to an extremely low (sub-solar) metallicity of $Z/Z_\odot\approx0.4$. This implies that the BLR metallicity for this group is nearly 40 times lower than that of the \siiv Excess quasars. Additionally, we also note that the median  \mgii integrated flux shows a substantial  increase of approximately 1.8 times compared to the median of \citet{Wu_2022} quasars.  

The black dash-dot spectrum in the bottom right panel of Fig.~\ref{fig:Anomaly_Clusters_and_Anomaly_Spectra} shows the mean spectra of the members of this group. It closely resembles the Flat BAL spectrum (see panel (d) in Fig.~\ref{fig:all anomaly spectra}), which is predominantly composed of LoBAL quasar spectra. This resemblance arises because, in BAL quasars, the wide absorption troughs also result in a reduction of \siiv emission line flux, creating a similar spectral appearance. Consequently, some of the \siiv Deficient anomalies appeared as minor members of Group 3 in the {\tt\string Full dataset}. However, the exclusion of BAL spectra allowed for the full identification of these anomalies in Group 3 of the {\tt \string Non-BAL Only Dataset}.

\subsection{BAL Anomalies} \label{discussion:bal anomalies}
There are four types of BAL anomalies as discussed in \S \ref{result:BALAnomalies}. The majority of HiBAL quasars are not identified as anomalies because HiBAL QSOs, which constitute approximately 85\% of the BAL population, are well captured by the PCA. The eigenvectors effectively adapt to map HiBAL features, ensuring that most typical HiBAL quasars are not marked as anomalies. As a result, the four BAL anomaly groups represent a minor subset of the BAL population, primarily due to the large number of HiBAL quasars that are included in the dataset.

\subsubsection{Blue BALs} \label{blue bals}
We define Blue BALs as BAL quasars with a significantly bluer continuum spectral index $(\alpha)$. These quasars exhibit strong \civ and \siiv absorption lines paired with a steep, downward-sloping spectrum resembling the \siiv excess anomalies (see panel (d) and (b) in Fig.~\ref{fig:all anomaly spectra}). They do not show any signs of reddening and have significantly higher emission flux towards the high-energy end of the spectrum, which rapidly decreases with increasing wavelength. This is unusual of a BAL quasar as most of them are known to be redder than a typical quasar \citep{menou2001broad}. Due to the absence of reddening, they are positioned at the bottom part (with low PCA 2 coefficients) in the left panel of Fig.~\ref{fig:BAL and Red Spread}. The \mgii emission line is feeble or visibly absent in most of these quasars. This category of BAL anomalies is rare with only 64 such quasars detected in our analysis, with all of them being HiBAL quasars. Interestingly, the spectral shape of these Blue BAL quasars coincide with the \citet{2001AJ....122..549V} composite, with the only difference being the deep \civ and \siiv absorption troughs. This can be seen as an exception to the results by \citet{reichard2003continuum} who found all HiBAL quasars in their study to be prominently redder than the EDR (SDSS Early Data Release) QSOs.

\subsubsection{Flat BALs}
The quasars in this group have a flat spectrum with nearly the same continuum flux throughout the wavelength range i.e. nearly a zero power law. This feature is indicative of significant reddening which diminishes the flux at shorter wavelengths, making it comparable to the longer wavelength  flux \citep[see][]{hopkins2004dust}. Their spectral slope is in between the Blue BALs (steep downward slope) and Reddened BALs (steep upward slope) which corresponds to their placement in the PCA eigenspace (the green scatter points in the left panel of Fig.~\ref{fig:BAL and Red Spread}). 

About 80\% of these flat BALs are LoBAL QSOs or \mgii BAL quasars. LoBALs are an important yet poorly understood class of quasars that provide direct evidence for energetic and variable mass outflows \citep[e.g., see][]{2014MNRAS.440..799V,yi2019variability,yi2019vizier}. As noted by \citet{wethers2019role}, LoBALs constitute about 15\% of the total BAL quasar population, with HiBAL quasars making up the majority. 
As seen in panel(d) of Fig.~\ref{fig:all anomaly spectra}, the composite Flat BAL spectrum (green) is significantly redder than both the Blue BAL composite (blue) and the \citet{2001AJ....122..549V} composite, consistent with the findings of \citet{reichard2003continuum}. This is the largest BAL anomaly group with a total of 306 members out of which 9 BALs are also cataloged by \citet{trump2006catalog}.

\subsubsection{Red BALs} \label{Reddened BALs}
These quasars exhibit a strongly positive-sloped spectrum with a redder continuum spectral index, indicating significant dust reddening (see red spectrum in panel (d) of Fig.\ref{fig:all anomaly spectra}), corresponding to significantly higher PCA 2 coefficients.
The members of this group have broad \civ absorption lines, which significantly absorb the blue part of the emission line.
Most of the members also have prominent iron absorption lines. This group consists of 213 objects, with 60\% being FeLoBALs and the remaining 40\% LoBAL quasars. The FeLoBAL quasars in this group differ from the classical FeLoBALs (another subgroup within BAL anomalies) primarily because their iron absorption lines are generally narrower. Due to this distinction and the occasional ambiguity in identifying \feii\ absorption caused by strong reddening, combined with low \feii\ strength,  we chose not to combine these FeLoBALs with the classical FeLoBALs. 

Notably, these BAL quasars are much redder as compared to the composite LoBAL spectra as created by \citet{reichard2003catalog}. 
\citet{2015MNRAS.453.3932R} study a population of extremely reddened BAL quasars, to analyse reddening as function of orientation and intrinsic nature of quasars. It is interesting to  note that the members of this group show significantly higher reddening than the Flat BAL group. Consistently, \citet{hamann2017extremely} confirmed that BALs quasars are more common with stronger or deeper profile in redder quasars.

\subsubsection{FeLoBALs}
The rare class of LoBAL quasars with absorptions from meta-stable excited states of \feii are known as FeLoBAL quasars \citep{becker2000properties,Choi2022}. This group consists exclusively of extreme cases of these BAL quasars. Typical members exhibit properties such as absorptions from over 20 transitions involving at least a dozen elements, similar to SDSS 1723+5553, as analyzed by \citet{hall2002unusual}. The broad-band spectral properties of FeLoBALs are thought to indicate scenarios such as mergers, star burst, etc. \citep[For e.g.][]{farrah2010extraordinary}. \enquote{Unusual} FeLoBALs can also be indicative of unique mechanisms such as a resonance-scattering interpretation of FeLoBAL 1214+2803 by \citet{branch2002spectrum}.

The composite spectrum of the 109 members of this group is shown in panel (d) of Fig.~\ref{fig:all anomaly spectra}. Most members feature extremely broad and deep absorption lines that significantly obscure the continuum, resulting in a highly unusual spectrum. In the PCA eigenspace, these quasars occupy a region similar to that of the Reddened BALs, as FeLoBALs represent some of the most extreme cases of reddening observed in BAL quasars.

\subsection{Reddened Anomalies} \label{reddened anomalies}
Reddened quasars are a subset of AGNs whose spectra exhibit significant attenuation at shorter wavelengths due to the presence of dust. This dust absorbs high-energy ultraviolet and blue light, re-emitting it at longer, redder wavelengths, producing a characteristic \enquote{reddening} effect. These quasars provide unique insights into the interplay between quasar activity and the surrounding interstellar medium, as well as the dust content within the host galaxy \citep[see][]{shao2022interstellar}. Reddened quasars are often associated with high levels of star formation and dust-rich environments, which may obscure the central black hole and impact the observed properties of the quasar \citep[e.g.][]{andonie2022panchromatic}.
As discussed by \citet{hopkins2004dust}, there could be several other processes too that result in a redder spectrum such as an intrinsically red continuum, an excess of synchrotron emission in red, intervening absorption by galaxies along the line of sight, or as discussed, dust extinction in the host galaxy or the quasar central engine itself. In our analysis, we identify three groups of reddened  anomalies which are discussed below.

\subsubsection{Heavily Reddened Quasars}
This group consists of quasars exhibiting a spectrum with excessive positively sloped continuum (see panel (e) in Fig.~\ref{fig:all anomaly spectra}) i.e. significantly redder continuum spectral index. Due to the reddening, they are represented by high values of PCA 2 coefficient placing them in the topmost region of the PCA eigenspace as shown by red scatter points in the right panel of Fig.~\ref{fig:BAL and Red Spread}.  
Nearly all members of this group closely resemble  the \enquote{too red} composite seen in \cite[Fig.~7][]{richards2003red}, indicating that this group represents the extreme cases of reddening in quasars. Additionally, some members also feature narrow absorption lines. \citet{hopkins2004dust} showed that reddened quasars are much more likely to show narrow absorption at the redshift of the quasar than are unreddened quasars.
\begin{table*}[ht]
\centering
\setlength{\tabcolsep}{10pt}  
\renewcommand{\arraystretch}{1}  

\begin{tabular*}{\textwidth}{lccccccl}
\hline
Name &RA  & Dec  & red- & r-band  & Cluster  & Group  & Anomaly \\ 
& (deg) & (deg) & shift & magnitude & ID & ID &  Type \\
\hline
\hline
SDSS J001439.18-045138.9 & 3.6632 & -4.8608 & 2.1261 & 21.19 & 2 & 1 & \civ Peaker \\ 
SDSS J102835.79+262923.4 & 157.1491 & 26.4898 & 2.0888 & 18.50 & 1 & 2 & Excess \siiv  \\ 
SDSS J002048.12-032444.4 & 5.2005 & -3.4123 & 1.9333 & 20.24 & 1 & 3 & \siiv Deficient \\ 
SDSS J091333.72+132541.3 & 138.3905 & 13.4281 & 2.1168 & 19.21 & 1 & 3 & Blue BAL \\ 
SDSS J000228.71+103732.8 & 0.6196 & 10.6257 & 1.9522 & 20.68 & 1 & 3 & Flat BAL \\ 
SDSS J091836.30+523711.0 & 139.6512 & 52.6197 & 1.9157 & 19.91 & 1 & 3 & Red BAL \\ 
SDSS J123015.99+062056.7 & 187.5666 & 6.3491 & 1.8512 & 20.05 & 1 & 3 & FeLoBAL \\ 
SDSS J125831.40+522124.6 & 194.6308 & 52.3568 & 1.9237 & 19.89 & 3 & 4 & Heavy Red \\ 
SDSS J133017.54+044646.4 & 202.5731 & 4.7795 & 2.2054 & 18.79 & 3 & 4 & Moderate Red \\ 
SDSS J124331.66+520335.1 & 190.8819 & 52.0597 & 2.1730 & 18.51 & 3 & 4 & Plateau Shaped \\ \hline
\end{tabular*}
\caption{Sample table listing the anomalous quasars identified in this work. A complete version of the table is available in machine-readable format.}
\label{table:sample_fulltable}
\end{table*}

\subsubsection{Moderately Reddened Quasars}
This group refers to quasars with either a rather constant flux throughout the wavelength range. Visual inspection revealed that this group actually consists of two sub-groups: one with a flat continuum and another with a convex-shaped continuum (see green plot in panel (e) of Fig.~\ref{fig:all anomaly spectra}). The reddening in this group is less pronounced compared to heavily reddened quasars.  The mean spectrum of this group traces the \enquote{dust reddened} composite of \cite[Fig.~7]{richards2003red}, which features a zero power law continuum between the rest wavelength of 1300\AA\ and 3000\AA. They are placed aptly below the heavily reddened quasars in the right panel of Fig.~\ref{fig:BAL and Red Spread}, with smaller PCA 2 coefficients representing the lesser extent of reddening. The \mgii emission in the members of this group is nearly identical to that of the heavily reddened quasars. This group also contains peculiar cases of quasars with a spectrum resembling an inverted parabola. We call these quasars as \enquote{convex} members. The inverted-parabola spectrum of these convex members has a peak luminosity around 2000 Å which tapers off symmetrically toward both ends. Some members also display narrow absorption lines, particularly toward the redder part of the spectrum. The spectra of these convex members is identical to that of the Plateau-shaped spectrum quasars between 2000 and 3000\AA. But, between 1250 and 2000\AA, where the Plateau-shaped spectrum quasars show a constant flux, convex members show a sharp positively sloped continuum. In terms of visual appearance, the convex quasars are closer to the Plateau-shaped spectrum quasars, but are placed with the Moderately reddened quasars instead. This is because, in the PCA projection (see the right panel of Fig.~\ref{fig:BAL and Red Spread}) the convex members are placed along with the Moderately reddened quasars (in the green group) whereas, the Plateau-shaped spectrum quasars are spatially separated and clustered distinctively.

\subsection{Plateau-shaped Spectrum Quasars} \label{humped quasars}
 The continuum of these quasars can be characterised into two parts. The first half (1250--2000\AA) exhibits a relatively flat continuum with broad \siiv and \civ emission lines. The second half (2000--3000\AA) rapidly falls with increasing wavelength. These quasars are termed as \enquote{Plateau-shaped Spectrum} quasars. This is because the flat (or slightly convex) feature together with the steeply sloped latter half of the spectrum imparts a plateau appearance to the continuum (see panel (f) in Fig.~\ref{fig:all anomaly spectra}). The members of this group have a nearly equally strong \civ and \siiv emission peak, which is caused by substantially less \civ emission flux as compared to a typical quasar. The \heii strength as well as the flux between 1640\AA\, and 1910\AA\, in these quasars is significantly more than that of a typical quasar. These quasars also show a strong \mgii emission line along with an enhanced iron emissions between 2250 and 2750\AA. 

\subsection{Machine Error Anomalies}
All the anomaly groups also consistently include anomalies caused by corrupted spectra. These anomalies often feature extended regions of missing flux data or partial spectra that are either padded with zeros or interpolated with a line connecting the disjoint sections. One of the most consistent anomalies is the quasars featuring a sharp narrow Dirac-Delta function-like peak caused by a cosmic ray encounter as discussed in \S \ref{subsec:CIV Peakers}. These were removed from the \civ Peakers group by a simple threshold cut on the \civ line equivalent width. Other occasional machine error anomalies were visually identified and discarded, as they do not contribute to any meaningful scientific analysis.

The full list of anomalous quasars identified in this work is available as a value-added catalog with this paper. Table.~\ref{table:sample_fulltable} provides a sample list, including the quasar name, Right Ascension, Declination, redshift, along with the cluster group numbers and their final anomaly classification.

\section{Conclusion} \label{sec:Conclusion}
Applying hierarchical K-Means clustering in a 20-dimensional PCA eigenvector hyperspace representing quasar spectra; We have presented five broad categories of quasar anomalies divided into 10 homogeneous groups i.e.
\begin{enumerate}
    \item \textbf{\civ Peakers:} A total of 65 quasars with extremely strong yet narrow (median $\sim2000$\kms) \civ emission line.
    
    \item \textbf{Excess \siiv emitters:} Quasars in this group exhibit an excessively high \siiv\ to \civ\ emission line flux ratio, nearly double the median value observed for all quasars in \citet{Wu_2022} catalog. A total of 227 such quasars were found.  
    
    \item \textbf{\siiv Deficient Anomalies:} 328 quasars with disproportionately low \siiv emission with \siiv to \civ flux ratio being one third of the median ratio for all quasars in \citet{Wu_2022} catalog.

    \item \textbf{BAL Anomalies:} A total of 692 quasars were identified as anomalies with BAL profiles. These were further subdivided into four sub-groups as follows:
    \begin{itemize}

    \item \textbf{Blue BALs:} A total of 64 HiBAL quasars with a strong negatively sloped (\enquote{blue}) continuum which is atypical of a BAL quasar. 
    
    \item \textbf{Flat BALs:} 306 BAL quasars with a relatively flat continuum. Among these, 80\% are LoBAL quasars 
    
    \item \textbf{Reddened BALs:} A total of 213 BAL quasars with heavily reddened continuum, hence a strong positively slope spectra. 
    
    \item \textbf{FeLoBALs:} These are a very rare class of BAL quasars with strong Fe absorptions and heavily reddened continua. There are a total of 109 such quasars detected. 
    \end{itemize}

    \item \textbf{Reddened Anomalies:}
    A total of 341 quasars were identified as reddened anomalous quasars, characterized by extreme reddening, as evident from their significantly red spectral slope.  They were further subdivided into three sub-groups based primarily on the degree of reddening as follows:
    \begin{itemize}

    \item \textbf{Heavily Reddened Quasars:} A total of 165 quasars were identified with steep, positively sloped spectra, attributed to heavy dust reddening.

    \item \textbf{Moderately Reddened Quasars:} Dust reddened quasars with relatively flat or slightly convex shaped continuum. A total of 93 such quasars were found.

    \item \textbf{Plateu-Shaped Spectrum Quasars:} A total of 83 peculiar quasars were identified with a plateau-shaped spectrum, characterized by a flat continuum followed by a negatively sloped continuum. These quasars exhibit a nearly flat continuum between 1250 and 2000 \AA, which then rapidly declines as the wavelength increases from 2000 to 3000 \AA.
      \end{itemize}
\end{enumerate}

This work has significantly expanded the number of sources in each anomalous quasar group. For example, we present a sample of 109 extreme FeLoBALs with strong, broad \feii\ absorption, along with an additional 127 FeLoBALs from the red BAL anomalies, where the iron absorption lines are considerably narrower. We have developed an efficient method for identifying anomalous quasars and classifying them into distinct categories.  The detected anomalies are presented in a value-added catalog, available with this paper. This approach has been successfully applied to the rest-frame UV spectra from the SDSS DR16 catalog and is currently being prepared for broader implementation. We applied the same algorithm to rest-frame optical spectra from SDSS DR16, covering around 75,000 quasars in the redshift range 
 $0.1\leq z\leq1.1$, including key emission lines such as {O~{\sc iii}}, H$\beta$, and optical iron. This catalog will be crucial, as these emission lines—particularly in the context of Eigenvector 1—are tightly correlated with the accretion rate and orientation of quasars \citep[see][]{shen2014diversity}. The catalog is currently under preparation. 
 
 With upcoming large-scale surveys such as Dark Energy Spectroscopic Instrument (DESI), 4-metre Multi-Object Spectroscopic Telescope (4MOST), and William Herschel Telescope (WHT) Enhanced Area Velocity Explorer (WEAVE), the number of quasar spectra will increase significantly, and our methodology will aid in identifying anomalous quasars in these vast datasets. Additionally, our approach can be adapted for photometric surveys like the Legacy Survey of Space and Time (LSST) survey to find photometric anomalous quasars.   By extending the sample size of these anomalous quasars, this study enables the statistical analysis of these peculiar sources, contributing to a deeper understanding of AGN and their diverse characteristics.

\begin{acknowledgements}
      AT acknowledges and thanks the Indian Institute of Astrophysics (IIA) for their acceptance in the visiting student program (VSP) and hence the financial and infrastructural support provided. MV acknowledges support from Department of Science and Technology, India - Science and Engineering Research Board (DST-SERB) in the form of a core research grant (CRG/2022/007884).

Funding for the Sloan Digital Sky Survey IV has been provided by the Alfred P. Sloan Foundation, the U.S. Department of Energy Office of Science, and the Participating Institutions. SDSS acknowledges support and resources from the Center for High-Performance Computing at the University of Utah. The SDSS web site is www.sdss4.org.

SDSS is managed by the Astrophysical Research Consortium for the Participating Institutions of the SDSS Collaboration including the Brazilian Participation Group, the Carnegie Institution for Science, Carnegie Mellon University, Center for Astrophysics | Harvard \& Smithsonian (CfA), the Chilean Participation Group, the French Participation Group, Instituto de Astrofísica de Canarias, The Johns Hopkins University, Kavli Institute for the Physics and Mathematics of the Universe (IPMU) / University of Tokyo, the Korean Participation Group, Lawrence Berkeley National Laboratory, Leibniz Institut für Astrophysik Potsdam (AIP), Max-Planck-Institut für Astronomie (MPIA Heidelberg), Max-Planck-Institut für Astrophysik (MPA Garching), Max-Planck-Institut für Extraterrestrische Physik (MPE), National Astronomical Observatories of China, New Mexico State University, New York University, University of Notre Dame, Observatório Nacional / MCTI, The Ohio State University, Pennsylvania State University, Shanghai Astronomical Observatory, United Kingdom Participation Group, Universidad Nacional Autónoma de México, University of Arizona, University of Colorado Boulder, University of Oxford, University of Portsmouth, University of Utah, University of Virginia, University of Washington, University of Wisconsin, Vanderbilt University, and Yale University.
\end{acknowledgements}

\begin{figure*}[h!]
    \centering
    \includegraphics[width=0.7\hsize]{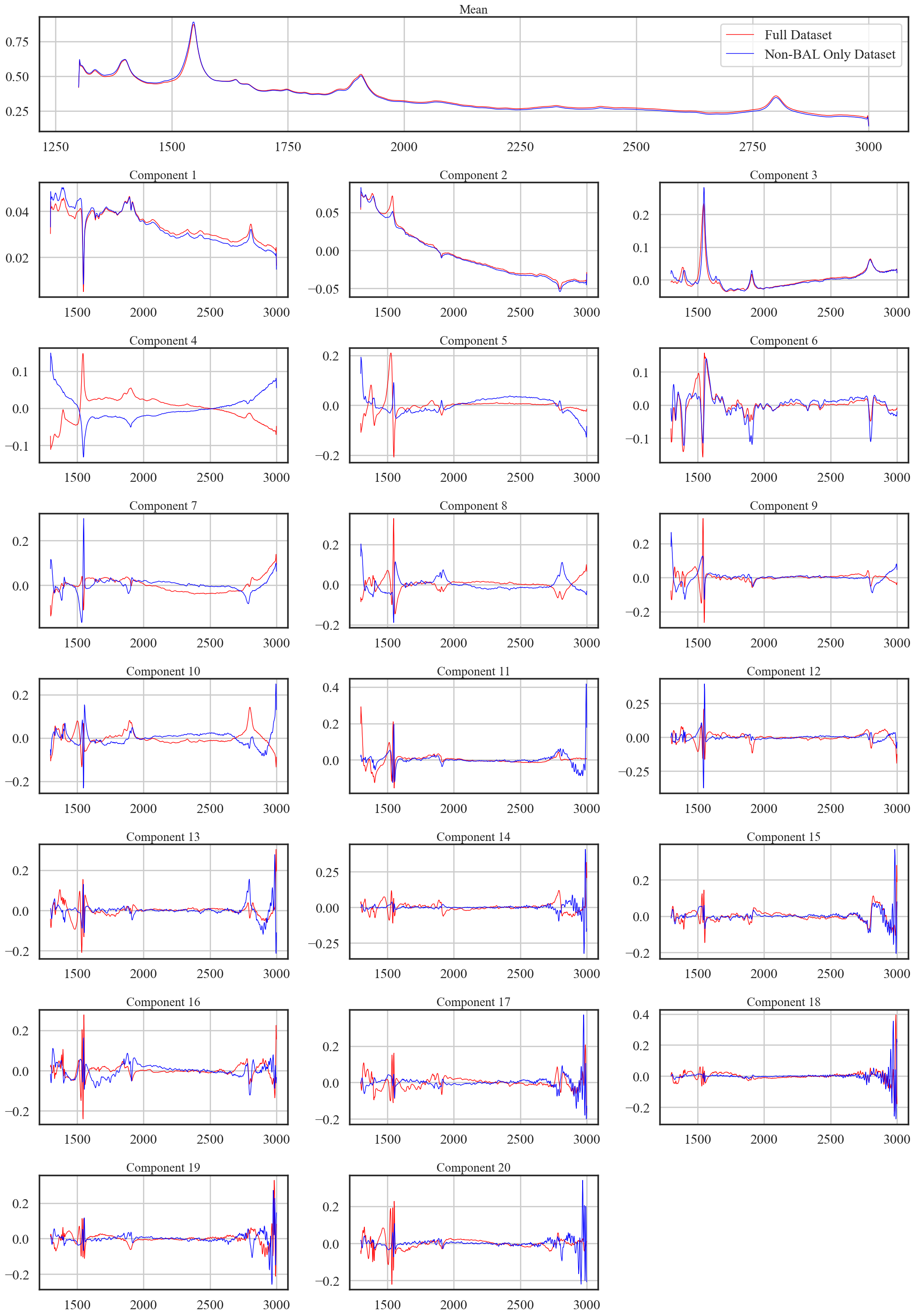}
    \caption{Each panel in the figure shows one of the 20 components of the PCA decomposition. Each component represents an eigenvector of the 20 component PCA. The first plot on the top depicts the mean spectrum of the total dataset. This mean spectrum is coincident with the \citet{2001AJ....122..549V} composite spectrum which reinforces the generalized nature of our selected data sample.}
    \label{fig:pca_eigenspectra}
\end{figure*}

%
%
\bibliographystyle{aa}
\bibliography{SQuAD1AA}
\appendix
    The PCA Eigenvectors for the primary K-Means clustering are given in Fig.~\ref{fig:pca_eigenspectra}. 
\end{document}